\documentclass{article}
\usepackage{tex/latex-template/arxiv/arxiv}
\usepackage[T1]{fontenc}
\usepackage{times}
\usepackage{multirow}
\usepackage{array}
\usepackage{makecell}
\usepackage[table]{xcolor}
\usepackage{siunitx}
\usepackage{caption}
\usepackage{float}
\usepackage{enumitem}
\graphicspath{{figs/}{tex/latex-template/arxiv/}}
\microtypesetup{expansion=false}
\DeclareRobustCommand{\texttt}[1]{#1}
\crefname{figure}{Fig.}{Figs.}
\Crefname{figure}{Figure}{Figures}
\crefname{table}{Table}{Tables}
\Crefname{table}{Table}{Tables}
\crefname{section}{Section}{Sections}
\Crefname{section}{Section}{Sections}
\newcommand{\figlink}[2]{\hyperref[#1]{Fig.~#2}}
\newcommand{\tablink}[2]{\hyperref[#1]{Table~#2}}
\newcommand{\seclink}[2]{\hyperref[#1]{Section~#2}}
\newcommand{\suppfiglink}[2]{\hyperref[#1]{Supplementary Fig.~#2}}
\newcommand{\supptablink}[2]{\hyperref[#1]{Supplementary Table~#2}}

\title{StretchCast: Global-Regional AI Weather Forecasting on Stretched Cubed-Sphere Mesh}
\renewcommand{\shorttitle}{StretchCast: global-regional AI forecasting on stretched cubed-sphere mesh}

\author[1,2]{Jin Feng}
\affil[1]{Institute of Urban Meteorology, China Meteorological Administration (IUM, CMA), Beijing, China}
\affil[2]{LeadSee Team\\Beijing Meteorological Service}
\date{}

\hypersetup{
  pdftitle={StretchCast: Global-Regional AI Weather Forecasting on Stretched Cubed-Sphere Mesh},
  pdfauthor={Jin Feng},
  pdfsubject={AI weather forecasting},
  pdfkeywords={AI weather forecasting, stretched cubed-sphere, global-regional forecasting},
  colorlinks=true,
  linkcolor=blue,
  citecolor=blue,
  urlcolor=blue,
}

\begin{document}
\maketitle
\begin{abstract}
Global AI weather forecasting still relies mainly on uniform-resolution models, making it hard to combine regional refinement, two-way regional-global coupling, and affordable training cost. We introduce \texttt{StretchCast}, a global-regional AI forecasting framework built on a variable-resolution stretched cubed-sphere (\texttt{SCS}) mesh that preserves a closed global domain while concentrating resolution over a target region. Within this framework, we develop a one-step predictor, \texttt{SCS\_Base Model}, and a rollout-oriented multistep predictor, \texttt{SCS\_FCST4 Model}, to test the feasibility of \texttt{SCS}-based forecasting and the benefit of joint multistep training. Experiments use \texttt{ERA5} with \texttt{69} variables over \texttt{1998-2022}. Because training compute remains limited, this study uses a coarse-resolution proof-of-concept configuration rather than a final high-resolution system. Even with only about \texttt{7,776} effective global grid cells and roughly \texttt{0.875 degree} resolution over the center-refined face, the \texttt{23M}-parameter \texttt{SCS\_Base Model} yields stable multivariate forecasts. With \texttt{83M} parameters and training cost on the order of hours, \texttt{SCS\_FCST4 Model} delivers competitive medium-range anomaly-correlation evolution over the target region after unified reprojection, especially for geopotential height, specific humidity, and part of the lower-tropospheric winds, while maintaining smooth cross-face continuity and realistic multiscale structure in typhoon and spectral analyses. These results support \texttt{StretchCast} as a practical lightweight foundation for global-regional AI weather forecasting.
\end{abstract}

\section{Introduction}\label{sec:intro}
Data-driven weather forecasting has become one of the fastest-moving areas of AI for Earth system science, yet most recent progress still centers on uniformly high-resolution global modeling. The horizontal mesh fundamentally determines the design route of an AI weather model. Existing global AI weather models roughly follow two mesh choices. The first uses regular latitude-longitude grids. \texttt{Pangu-Weather}, \texttt{FengWu}, and \texttt{FuXi} have shown that regular rectangular tensor backbones are well suited to global medium-range weather forecasting because dense convolution and attention operators align well with modern accelerators and because parameter sharing is natural on image-like arrays \citep{bi2023pangu,ren2023fengwu,chen2023fuxi}. For region-focused forecasting, however, this route has an immediate limitation: if the globe remains uniformly high resolution, then giving the target region finer representation also forces the model to pay the same memory and training cost over all other regions. At the same time, polar singularity cannot be avoided in that representation. The second route is represented by \texttt{GraphCast}, which uses graph structures that more closely follow spherical topology and learns directly on nearly uniform spherical discretizations, thereby avoiding latitude-longitude singularities more directly \citep{lam2023graphcast}. This choice is geometrically attractive, but it is also more expensive to train and run. For multivariate, multilayer forecasting systems that require long rollouts, irregular message passing does not align with efficient dense operators as naturally as regular tensor blocks do. In short, existing mesh designs usually capture either tensor-computation efficiency or spherical geometric consistency, but not both at once.

Operational weather forecasting needs, meanwhile, are often regional. Forecasters care about local heavy rainfall, cold-air outbreaks, and typhoon events. For such problems, the main question is not whether a data-driven forecasting model should spend the same amount of computation everywhere on the globe. The more practical question is whether dense computation can be allocated to regions of interest, sparse computation to the rest of the globe, and global and regional evolution kept tightly coupled at the same time.

We therefore seek an AI forecasting route better suited to integrated global-regional modeling: one that preserves sparse background computation over a full global domain, allocates higher resolution and denser computation to the target region, reuses regular rectangular tensor operations as much as possible to reduce training cost, avoids polar singularity, and remains competitive in forecasting performance. A natural candidate is the cubed-sphere mesh, which has long been used in numerical weather prediction (NWP). It discretizes the sphere into six connected rectangular faces with consistent topology. In doing so, it avoids meridian convergence near the poles, reduces high-latitude distortion, removes pole singularity, and still preserves a regular rectangular topology convenient for modeling. Long practice in NWP also shows that this mesh family is not only geometrically sensible but operationally stable. \citet{putman2007finite} established the feasibility of cubed-sphere transport on the globe, and the \texttt{FV3} family later turned the cubed sphere into a mature dynamical-core framework. The stretched extension of the cubed sphere goes one step further by showing that regional refinement can be introduced within a closed global domain while still preserving two-way coupling with the global background and the large-scale circulation constraints that affect the target region \citep{harris2013twoway,harris2016highresolution}. In other words, the stretched cubed sphere is well matched to the requirement of dense regional computation plus sparse global background.

Based on that idea, we propose \texttt{StretchCast}. It brings the stretched cubed-sphere representation from NWP into an AI weather forecasting framework, builds a variable-resolution mesh centered over a target region, and instantiates it here over eastern China. On top of that mesh, we place two lightweight predictors: \texttt{SCS\_Base Model} and \texttt{SCS\_FCST4 Model}. The first tests whether this mesh can support meaningful multivariate weather prediction. The second tests whether rollout-oriented training can strengthen the same route at medium-range lead times. The same logic could also be applied to other target regions or reduced to a non-stretched cubed-sphere mesh. The question addressed here is more basic: under limited computational resources, can variable-resolution global-regional modeling on an \texttt{SCS} mesh become a viable data-driven weather forecasting route?

This work makes three contributions. First, we develop a medium-range data-driven weather forecasting model that uses a variable-resolution mesh for integrated global-regional prediction. Second, we adopt a neural backbone that fits modern tensor-parallel computation and enables an unusually lightweight training regime. This matters not only because it saves resources, but also because it shows that a regionally refined representation can already produce useful forecasting skill under a parameter budget far smaller than that of uniform high-resolution global models. Third, compared with much heavier globally uniform baselines, this lightweight design makes simultaneous multistep prediction tractable because graphics processing unit (\texttt{GPU}) memory can accommodate both the model and multistep training data, which in turn helps suppress rollout error accumulation at medium-range lead times.

\section{Related Works}\label{sec:related}
The work most directly related to this paper spans four directions: globally uniform AI forecasting, spherical graph modeling, variable-resolution meshes in NWP, and recent global-regional data-driven approaches. The most mature family of AI weather forecasting methods still operates on globally uniform latitude-longitude grids, represented by \texttt{Pangu-Weather}, \texttt{FengWu}, and \texttt{FuXi} \citep{bi2023pangu,ren2023fengwu,chen2023fuxi}. These methods typically use 3D convolutions, vision Transformers, or multistage cascade architectures to learn atmospheric evolution on regular tensors. \texttt{Aurora} extends the same globally uniform representation into an Earth system foundation-model setting \citep{bodnar2025aurora}. Another line of work, represented by \texttt{GraphCast}, uses graph structures on nearly uniform spherical discretizations to avoid latitude-longitude singularities more directly \citep{lam2023graphcast}. These two routes correspond to two dominant priorities in current AI weather modeling: regular-tensor efficiency and spherical geometric consistency. Both, however, mainly target globally uniform representations.

By contrast, NWP began discussing regional refinement inside a closed global domain much earlier. The cubed-sphere mesh has long been used to alleviate the geometric degeneration of latitude-longitude grids near the poles while providing more regular local adjacency on the sphere \citep{putman2007finite}. High-order nonhydrostatic dynamical cores based on multi-moment constrained finite volume (\texttt{MCV}) formulations have continued to use the same coordinate system \citep{li2020mcv}. More importantly, work on stretched grids, two-way nested cubed spheres, and variable-resolution global models has shown that regional refinement succeeds not only because the target region is finer, but also because the large-scale background outside the target region is still preserved and information can propagate in both directions between the two \citep{foxrabinovitz2008stretched,harris2013twoway,harris2016highresolution}.

More recent data-driven approaches have moved closer to the global-regional problem itself. \texttt{OneForecast} uses multi-scale graphs and nested regionally refined graph structures to place the global background and regional detail inside one graph representation \citep{gao2025oneforecast}. \texttt{YingLong} adopts a \texttt{3 km}, \texttt{1 h} limited-area AI model that depends on global AI forecasts for lateral boundary conditions \citep{xu2025yinglong}, and is therefore closer to high-resolution limited-area forecasting. \texttt{SR-Weather} and diffusion-based probabilistic downscaling methods mainly act as coarse-to-fine post-processing enhancement \citep{park2026srweather,molinaro2026diffusion}, rather than learning atmospheric evolution directly from a variable-resolution global representation. Taken together, current work covers globally uniform tensor modeling, spherical graph modeling, variable-resolution spherical meshes, and regional post-processing enhancement, but still lacks a lightweight global-regional AI forecasting route that combines the stretched cubed-sphere representation with efficient tensor-based neural backbones.

\section{Data and Methods}\label{sec:data-methods}

\subsection{Data and training objective}\label{sec:data-objective}
As in many other data-driven medium-range forecasting studies, the experiments here use \texttt{ERA5} global reanalysis on a \texttt{0.25 degree} latitude-longitude grid \citep{hersbach2020era5}. The input consists of \texttt{69} channels: five 3D atmospheric variables, geopotential height (\texttt{Z}), specific humidity (\texttt{Q}), temperature (\texttt{T}), zonal wind (\texttt{U}), and meridional wind (\texttt{V}), each expanded over \texttt{13} standard pressure levels from \texttt{1000-50 hPa}, together with four near-surface variables, \texttt{10 m} zonal wind (\texttt{U10m}), \texttt{10 m} meridional wind (\texttt{V10m}), \texttt{2 m} temperature (\texttt{T2m}), and sea-level pressure (\texttt{SLP}). All variables are sampled at \texttt{6 h} intervals. Because of training and storage constraints, we use \texttt{25} years of data from \texttt{1998-2022}, with \texttt{1998-2020} for training, \texttt{2021} for validation, and \texttt{2022} for testing. All main results reported below use the \texttt{2022} test set. Before entering the model, all inputs are remapped onto the \texttt{SCS} mesh used in this experiment, whose center-refined face is centered over eastern China.

The training objective is a channel-weighted mean squared error defined on the stretched cubed-sphere mesh. The channel weights follow the variable-weighting scheme used in \texttt{Pangu-Weather} \citep{bi2023pangu}. Classical latitude-longitude models often add latitude or area weights to make spherical error statistics area-consistent and to reduce the bias induced by denser high-latitude grids. We do not add such spatial weights here. The reason is not that area consistency is unimportant, but that this is a global-regional model focused on the central target region. Because the \texttt{SCS} mesh is denser over the central face and coarser in remote regions, removing additional area weighting means that the loss function naturally emphasizes the center-refined region. In other words, once the \texttt{SCS} mesh has already encoded regional priority, we do not want area weighting to flatten that priority again.

\subsection{SCS mesh as a global-regional representation}\label{sec:scs-mesh}
The cubed-sphere mesh discretizes the sphere into six connected spherical faces. The basic idea is straightforward: a cube is first used to approximate the sphere, and its six faces are then mapped onto the sphere, producing six faces with identical topology. Compared with the traditional latitude-longitude grid, this design avoids grid degeneration caused by meridian convergence near the poles, yields more regular local adjacency, and reduces the geometric distortion typical of spherical discretization at high latitude. For that reason, the cubed sphere has long been a widely used spherical mesh in global NWP and climate models \citep{putman2007finite}. The stretched cubed sphere introduces nonuniform resolution on top of that geometry. Its aim is not to keep the same spatial scale everywhere on the globe, but to redistribute grid points continuously toward a target region while still preserving a closed global integration domain, leaving regions farther away at relatively coarse resolution. The result is not a cut-out local nest but a variable-resolution mesh over the full sphere. Previous work on stretched grids and variable-resolution global models shows that this design can concentrate more resolution over priority regions without increasing the total computational burden too aggressively, while still preserving the large-scale circulation background and remote constraints that affect the target region \citep{foxrabinovitz2008stretched,harris2016highresolution}.

\begin{figure}[H]
\centering
\includegraphics[width=0.92\textwidth]{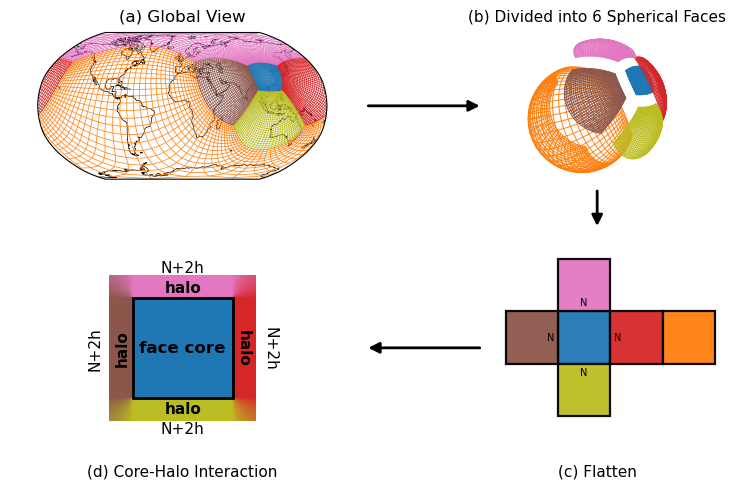}
\caption{Schematic of the stretched cubed-sphere (\texttt{SCS}) configuration used in this study and its \texttt{core-halo} interaction mechanism, with the center-refined face placed over eastern China. (a) Global view of the \texttt{SCS} grid and the center-refined region; (b) three-dimensional view of the sphere partitioned into six faces; (c) cross-shaped flattened arrangement of the six faces; and (d) \texttt{core-halo} structure of a single face, where the blue block denotes the face core, the surrounding colored bands denote halos, and \texttt{N} and \texttt{N+2h} indicate the corresponding numbers of grid points along the edges. Black arrows indicate the mapping from the spherical grid to the faceted representation and then to the flattened tensor representation.}
\label{fig:fig1}
\end{figure}

The \texttt{SCS} mesh used here is centered over eastern China. \figlink{fig:fig1}{1}(a) makes this resolution hierarchy explicit. In the discrete representation, the six spherical faces all have identical array size, and each face contains \texttt{36 x 36} grid cells, so the full globe has only \texttt{7,776} cells in total. After the stretching transformation, however, the physical resolution on the sphere differs by face. The center-refined face roughly covers \texttt{15.6-47.9 N, 90.7-130.9 E}, with local angular spacing of about \texttt{0.67-1.11 degree}, which forms the high-resolution core of the present configuration. The opposite remote face lies over the Southern Hemisphere, where local angular spacing can become as coarse as about \texttt{9 degree}, or roughly \texttt{1000 km}, and mainly serves to preserve global closure on the opposite half of the sphere. The four transition faces connect the central region smoothly to the background, with resolution near their centers around \texttt{2 degree} and an overall range of about \texttt{0.67-3.12 degree}.

The last three panels of \figlink{fig:fig1}{1} explain why this spherical mesh is suitable for deep learning. The six faces on the sphere can be split explicitly into surface patches, then flattened into equally sized 2D rectangular blocks arranged in a cross-shaped topology. This flattening lets all faces share the same 2D tensor backbone while keeping their topological relation explicit. A regular representation, however, does not mean that the six faces can be processed as independent image tiles. The \texttt{core-halo} design in the figure shows that, before each layer of local convolution or token mixing begins, the model explicitly injects boundary-neighbor information into each face through halo regions. In that way, geometric continuity on the sphere can be approximated while computation still runs on regular 2D tensors.

\begin{figure}[H]
\centering
\includegraphics[width=\textwidth]{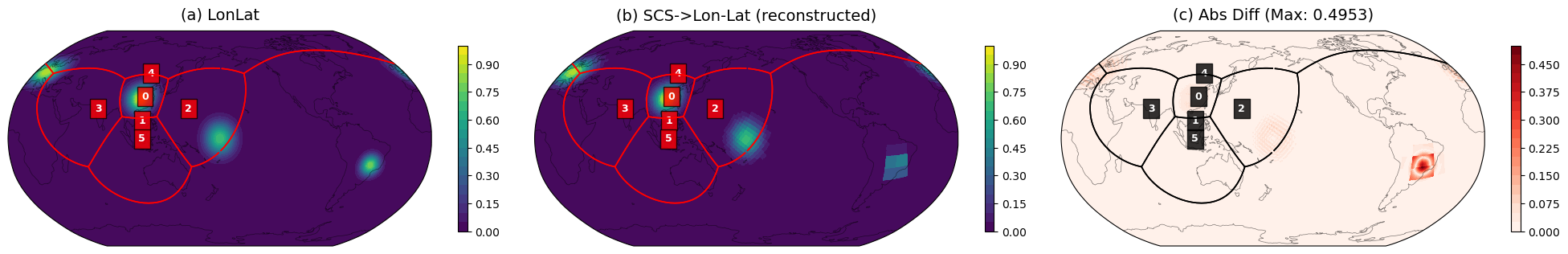}
\caption{Comparison between an idealized latitude-longitude field and its reconstruction from the \texttt{SCS} representation. (a) Original latitude-longitude field; (b) field reconstructed back from the \texttt{SCS} representation, with red curves marking face boundaries and red numbered boxes indicating face indices; and (c) absolute reconstruction error, with black curves marking face boundaries, a color bar showing the error magnitude, and the maximum absolute error reported in the panel title.}
\label{fig:fig2}
\end{figure}

\figlink{fig:fig2}{2} shows the reconstruction error produced when an idealized latitude-longitude field is mapped to \texttt{SCS} and then reconstructed. The test field is built from a set of large-value patches distributed across different faces and face intersections. The reconstruction shows no obvious visual seams near face boundaries, which means that the data mapping procedure used here is working as intended. \suppfiglink{fig:s1}{S1} extends the same diagnosis to a real \texttt{T2m} field and reaches the same conclusion. For a variable-resolution spherical mesh, however, mapping error cannot be spatially uniform across the whole globe: the smallest errors occur mainly over the center-refined region, while regions farther from the center and represented more coarsely show larger mapping errors (\figlink{fig:fig2}{2}(c), \suppfiglink{fig:s1}{S1}(c)). For models that must perform long rollouts, this raises a practical concern: can errors generated in remote regions propagate toward the central region through cross-face interaction and large-scale circulation pathways, causing the target region to lose predictability quickly as well? The results below show that this does not become the dominant problem. Although errors in remote regions grow faster, the center-refined region still remains stable during multistep rollout.

\subsection{Model design}\label{sec:model-design}
\texttt{StretchCast} uses two predictors. \texttt{SCS\_Base Model} is a one-step model that maps the atmospheric state at the current time to the state \texttt{6 h} later. \texttt{SCS\_FCST4 Model} takes two historical input frames and predicts four future frames at once, i.e., a \texttt{24 h} forecast segment. The first tests whether the \texttt{SCS} representation itself is already effective. The second is designed for rollout forecasting and asks whether multistep training can support more stable medium-range evolution.

\begin{figure}[H]
\centering
\includegraphics[width=\textwidth]{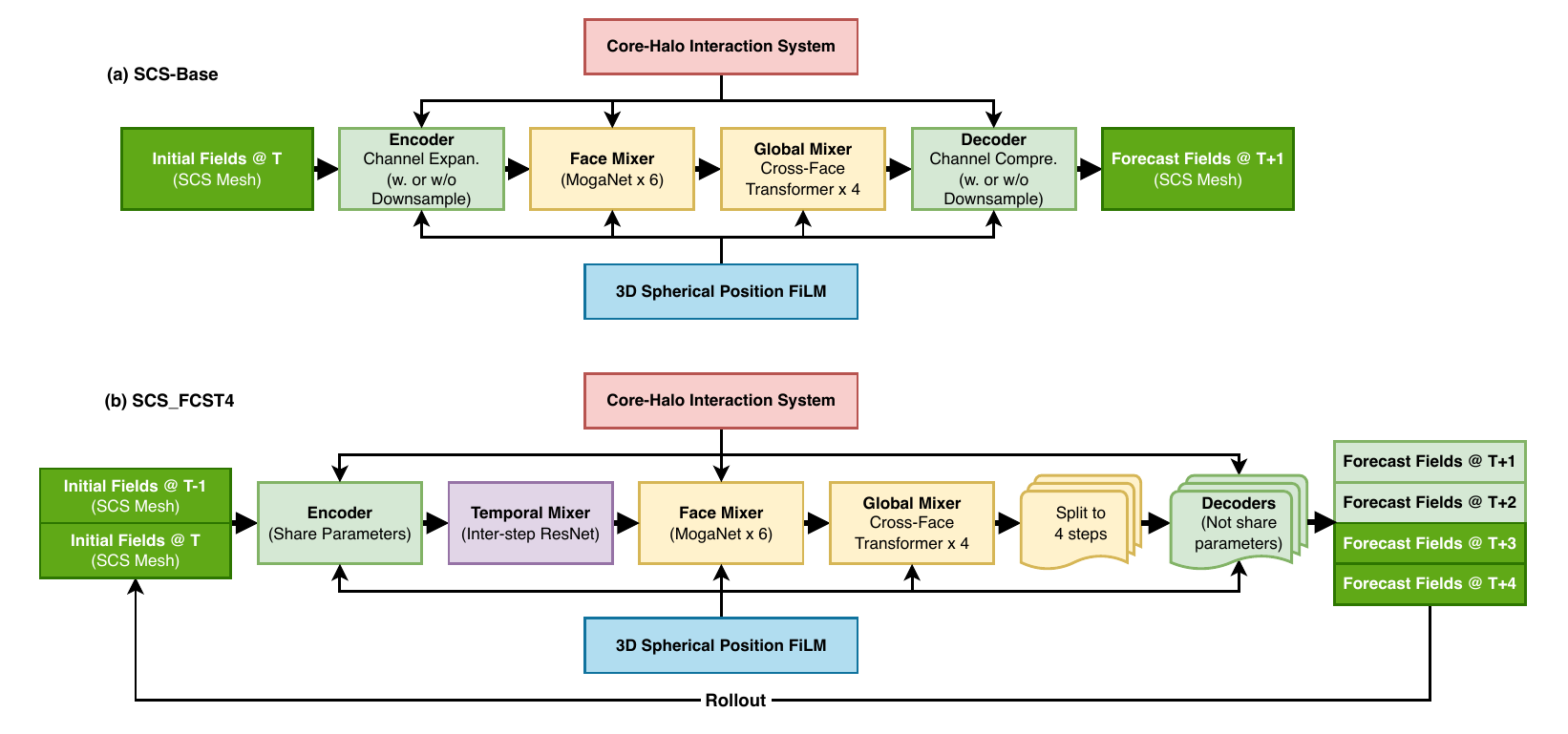}
\caption{Architecture of \texttt{StretchCast}. (a) Pipeline of the \texttt{SCS\_Base} one-step model. \texttt{Initial Fields @ T (SCS Mesh)} are processed by an \texttt{Encoder}, a \texttt{Face Mixer}, a \texttt{Global Mixer}, and a \texttt{Decoder} to produce \texttt{Forecast Fields @ T+1 (SCS Mesh)}. Pink modules denote the \texttt{Core-Halo Interaction System}, and blue modules denote \texttt{3D Spherical Position FiLM}. (b) Pipeline of the \texttt{SCS\_FCST4} multi-step model. \texttt{Initial Fields @ T-1} and \texttt{Initial Fields @ T} are encoded by a shared \texttt{Encoder}, processed by a \texttt{Temporal Mixer} and the same spatial backbone, and then passed to \texttt{Split to 4 steps} and multiple \texttt{Decoders} to output \texttt{Forecast Fields @ T+1}, \texttt{T+2}, \texttt{T+3}, and \texttt{T+4}.}
\label{fig:fig3}
\end{figure}

\figlink{fig:fig3}{3} gives the overall structure of \texttt{StretchCast}. This decomposition addresses four core modeling requirements: extracting local multiscale structure within each face, maintaining continuity across face boundaries, propagating global teleconnections, and handling temporal evolution. Inside each face, the model adopts the \texttt{MogaNet} token mixer previously validated in LeadSee-Precip to model local multiscale structure at relatively low cost \citep{ji2024leadsee,li2024moganet}. For interaction across faces, it splits neighboring boundary coupling and long-range cross-face dependency into two complementary paths, \texttt{core-halo} and \texttt{Global Mixer}, so that a single mechanism does not need to handle both local continuity and global information propagation.

That decomposition has a clear physical motivation. On a regionally refined spherical mesh, the evolution of local weather systems depends first on multiscale structure and short-range transport within a face, which requires a shared intra-face mixer. Yet weather fields do not stop at face boundaries. Thermodynamic and dynamical information must remain continuous across adjacent faces. To achieve that, the model uses the \texttt{core-halo} mechanism to re-inject neighboring boundary context after each layer of local mixing, thereby maintaining geometric continuity on the sphere. At the same time, boundary exchange alone is insufficient to capture teleconnections and large-scale background circulation constraints, so \texttt{Global Mixer} performs explicit patch-level cross-face interaction. The blue branch in the figure, \texttt{3D Spherical Position FiLM}, provides geometric conditional modulation. It follows the basic affine-modulation idea of \texttt{FiLM} \citep{perez2018film}, but encodes static geometric information such as local coverage, spherical center coordinates, and face identity into layerwise modulation signals (see \suppfiglink{fig:s2}{S2}), allowing spatial mixing to respond explicitly to the nonuniform geometry of the \texttt{SCS} mesh.

The one-step \texttt{SCS\_Base Model} can be written as a combination of an encoder, a spatial-mixing backbone, and a decoder. Let the \texttt{SCS} input at time \texttt{t} be $x_t \in \mathbb{R}^{C \times 6 \times H \times W}$, and denote the encoder and decoder by $E(\cdot)$ and $D(\cdot)$. The backbone forward process can then be summarized as

\[
z_t = E(x_t), \qquad \hat{x}_{t+1} = D\!\left(\mathcal{G}\!\left(\mathcal{F}(z_t)\right)\right),
\]

where $\mathcal{F}$ and $\mathcal{G}$ denote \texttt{Face Mixer} and \texttt{Global Mixer} in latent space, respectively. This form explicitly separates local intra-face modeling from global cross-face interaction: the former mainly organizes multiscale structure within a face, while the latter supplements the long-range dependencies that local propagation alone cannot recover.

For the $l$-th face-mixing layer, the model first performs shared-parameter token mixing on each face, and then generates FiLM modulation coefficients for that layer from static geometric fields. If \texttt{coverage}, spherical center coordinates, and face identity are denoted by $c$, $p$, and $e_f$, respectively, then

\[
\left[\gamma_l,\beta_l\right]
= \phi_l\!\left(\mathcal{H}_{\mathrm{add}}\!\left([c,p,e_f]\right)\right), \qquad
\tilde{z}_l
= \left(1+\gamma_l\right) \odot \mathcal{M}_l\!\left(\mathcal{H}_{\mathrm{add}}(z_{l-1})\right) + \beta_l,
\]

where $\mathcal{H}_{\mathrm{add}}$ denotes the operation that adds halo context to each face, $\mathcal{M}_l$ is the shared intra-face token mixer, and $\phi_l$ is a learnable mapping from static geometric maps to FiLM coefficients. Unlike ordinary positional encoding, the geometric information here is not injected only once at the input. It continuously modulates latent features layer by layer. This design allows the model to adapt feature transformation to local stretching degree and absolute spherical position, so that geometric nonuniformity participates directly in the construction of the forecasting operator.

The model then performs explicit cross-face updating over halo regions:

\[
z_l
= \mathcal{H}_{\mathrm{upd}}(\tilde{z}_l)
= \tilde{z}_l + \alpha \odot
\left[
\mathcal{H}_{\mathrm{add}}\!\left(\mathcal{H}_{\mathrm{rem}}(\tilde{z}_l)\right) - \tilde{z}_l
\right],
\]

where $\mathcal{H}_{\mathrm{rem}}$ removes halo regions and extracts the face core, and $\alpha$ is the boundary-mixing weight. The key point is not that cross-face interaction is patched in passively at the end. Rather, neighboring boundary context is re-injected in a geometrically consistent way after every layer of local mixing. For weather fields, this corresponds to continuously maintaining thermodynamic and dynamical continuity across boundaries throughout the network instead of treating the six faces as independent image tiles.

By contrast, \texttt{Global Mixer} targets the long-range dependencies that cannot be captured through boundary propagation alone. It first generates a global modulation signal from static geometric fields and face embeddings, then converts the latent features of the six faces into a patch sequence and feeds it to a Transformer:

\[
\left[\gamma_g,\beta_g\right] = \psi([c,p,e_f]), \qquad
T = \operatorname{Patch}\!\left(\left(1+\gamma_g\right)\odot z_t + \beta_g\right), \qquad
z_t^{\,\mathrm{glob}} = z_t + \eta \, \operatorname{Unpatch}\!\left(\operatorname{Trans}(T)\right).
\]

The model therefore separates "local cross-boundary continuity" and "global cross-face teleconnection modeling" into two complementary mechanisms: the former handles neighboring exchange and geometric continuity, while the latter handles large-scale background flow, teleconnections, and cross-region coupling that cannot be recovered by finite-depth boundary propagation.

Autoregressive rollout based on a one-step predictor will eventually accumulate error. For globally uniform \texttt{0.25 degree} latitude-longitude models, high-resolution global grids bring large input-tensor and activation cost, so training often optimizes only one-step prediction, with longer lead times obtained autoregressively at inference. That creates a mismatch between the training objective and rollout behavior, allowing errors to amplify step by step. Existing work uses different strategies to mitigate this problem. \texttt{FuXi} adopts subsequent autoregressive joint fine-tuning, \texttt{Pangu-Weather} uses multistep modeling with autoregressive inference, and \texttt{FengWu} uses replay-buffer-based self-corrective training \citep{bi2023pangu,chen2023fuxi,ren2023fengwu}. These strategies can reduce rollout drift to some extent, but often come with higher training cost, discontinuity between temporal blocks, or a tendency for forecast fields to shrink toward the mean state. \texttt{SCS\_FCST4 Model} adopts a different design: it first encodes historical frames into a shared latent space, and then performs explicit temporal integration:

\[
Z_t = \left[E(x_{t-K+1}), \ldots, E(x_t)\right], \qquad
\bar{z}_t = \mathcal{S}\!\left(\mathcal{T}(Z_t)\right), \qquad
\hat{x}_{t+1:t+4} = D(\bar{z}_t),
\]

where $\mathcal{T}$ denotes the temporal mixer and $\mathcal{S} = \mathcal{G} \circ \mathcal{F}$ denotes the spatial backbone applied afterward. For rollout forecasting, \texttt{SCS\_FCST4 Model} has three key design choices. First, it explicitly integrates the states of multiple time steps in latent space, and then lets the same spatial backbone propagate that information to local, cross-face, and global scales. The temporal mixer handles temporal information, while the spatial backbone projects that information into local weather structure, boundary exchange, and global circulation constraints. This is analogous to the basic logic of NWP, in which the state is advanced in time and then updated through physical processes. Second, because the \texttt{SCS} variable-resolution representation keeps both model size and activation cost relatively small, rollout-oriented training is directly feasible in the present setup. \texttt{SCS\_FCST4 Model} does not repeatedly apply the one-step predictor. Instead, it takes two historical input times and outputs four future steps at once. The benefit is not only higher throughput, but also fewer autoregressive applications and a shorter error-propagation chain in the training objective itself. If computational resources increase in the future, the same route can naturally be extended to longer multistep forecast segments. Third, the model adopts an asymmetric encoder-decoder design: the input side shares an encoder, while different lead times use different decoding branches, allowing different forecast horizons to learn their own corrections and projection mappings.

\subsection{Evaluation methods and comparison models}\label{sec:eval}
We evaluate both the center-refined face and the full global domain composed of all six faces. Results over the center-refined face correspond to the main application scenario of this study and most directly answer whether regional refinement brings stable benefit. Global results are used to test whether this regional refinement damages the consistency of the overall circulation. For \texttt{SCS\_Base Model}, simple autoregressive rollout is used only as a robustness check rather than as the recommended final long-range forecasting mode; for \texttt{SCS\_FCST4 Model}, multistep rollout forms the main basis of the long-range analysis throughout the paper.

The main external comparison models are \texttt{Pangu-Weather}, \texttt{FengWu}, and \texttt{FuXi}. This is not an arbitrary choice: they are the closest to the \texttt{69}-variable configuration used here. \texttt{Pangu-Weather} and \texttt{FengWu} both use \texttt{69} variables, while \texttt{FuXi} uses \texttt{70} input variables. Their inference products are obtained through NVIDIA Earth2Studio \citep{earth2studio_docs}, and then uniformly reprojected onto the same \texttt{SCS} mesh used in this paper so that differences caused by mesh and variable configuration are stripped away as much as possible. Because the original resolution, mesh system, and training setup are not exactly the same, the comparisons below are better interpreted as representation-level comparisons under unified reprojection rather than as a strict same-condition benchmark race.

\tablink{tab:atm_model_compute_cost}{1} places our models and representative global AI weather models in the same coordinate system of model size and training cost. Compared with globally uniform \texttt{0.25 degree} models, the center-refined \texttt{SCS} mesh used here requires only about \texttt{7,776} effective grid cells. \texttt{SCS\_Base Model} and \texttt{SCS\_FCST4 Model} have only \texttt{23M} and \texttt{83M} parameters, respectively, far below the \texttt{512M} of \texttt{Pangu-Weather} (combined across the \texttt{6 h} and \texttt{24 h} models), the \texttt{427M} of \texttt{FengWu}, and the \texttt{4,670M} of larger systems such as \texttt{Aurora} and cascade \texttt{FuXi}. In training cost, the main model in this paper requires only \texttt{8 x Hygon K100AI} GPUs, roughly equivalent to \texttt{3 x Nvidia A100}, and trains in about \texttt{18} hours. By contrast, representative uniformly high-resolution global models typically require tens to hundreds of high-end GPUs and training cycles measured in weeks.


\begin{table*}[t]
\raggedright
\scriptsize
\setlength{\tabcolsep}{1.8pt}
\renewcommand{\arraystretch}{1.06}
\resizebox{\textwidth}{!}{%
\begin{tabular}{@{}>{\raggedright\arraybackslash}p{0.13\textwidth}>{\raggedright\arraybackslash}p{0.09\textwidth}>{\raggedright\arraybackslash}p{0.09\textwidth}>{\raggedright\arraybackslash}p{0.11\textwidth}>{\raggedright\arraybackslash}p{0.08\textwidth}>{\raggedright\arraybackslash}p{0.08\textwidth}>{\raggedright\arraybackslash}p{0.18\textwidth}>{\raggedright\arraybackslash}p{0.10\textwidth}@{}}
\toprule
Organization & Model & Resolution & Num. grids & Meteo vars. & Params. & Computation resources for training & Time for training \\
\midrule
Huawei & Pangu & \multirow{4}{=}{0.25 (Global)} & \multirow{4}{=}{1110K} & 69 & 0.256B & Nvidia V100$\times$192 & 2.3 weeks \\
Shanghai Academy of AI for Science & FuXi &  &  & 70 & 4.69B & Not reported & Not reported \\
Shanghai AI Laboratory & FengWu &  &  & 69 & 0.427B & Nvidia A100$\times$32 & 17 days \\
Microsoft & Aurora &  &  & 84 & 1.3B & Nvidia A100$\times$32 & 2.5 weeks \\
ECMWF & AIFS & 0.25$^\circ$ (Reduced Gaussian) & 1110K & 90 state vars & 0.229B & Nvidia A100$\times$64 & 1 week \\
\midrule
IUM, CMA & StretchCast & 0.875$^\circ$ Center Face & 7.7K & 69 & 0.083B & Hygon K100AI DCUs ($\sim$35\% A100 computing power)$\times$8 & 18 hours \\
\bottomrule
\end{tabular}
}
\caption{Comparison of training compute resources, number of parameters, and training time across different atmospheric models.}
\label{tab:atm_model_compute_cost}
\end{table*}

\section{Results in SCS\_Base}\label{sec:results-base}

\subsection{Averaged performances}\label{sec:results-base-avg}
\begin{figure}[H]
\centering
\includegraphics[width=\textwidth]{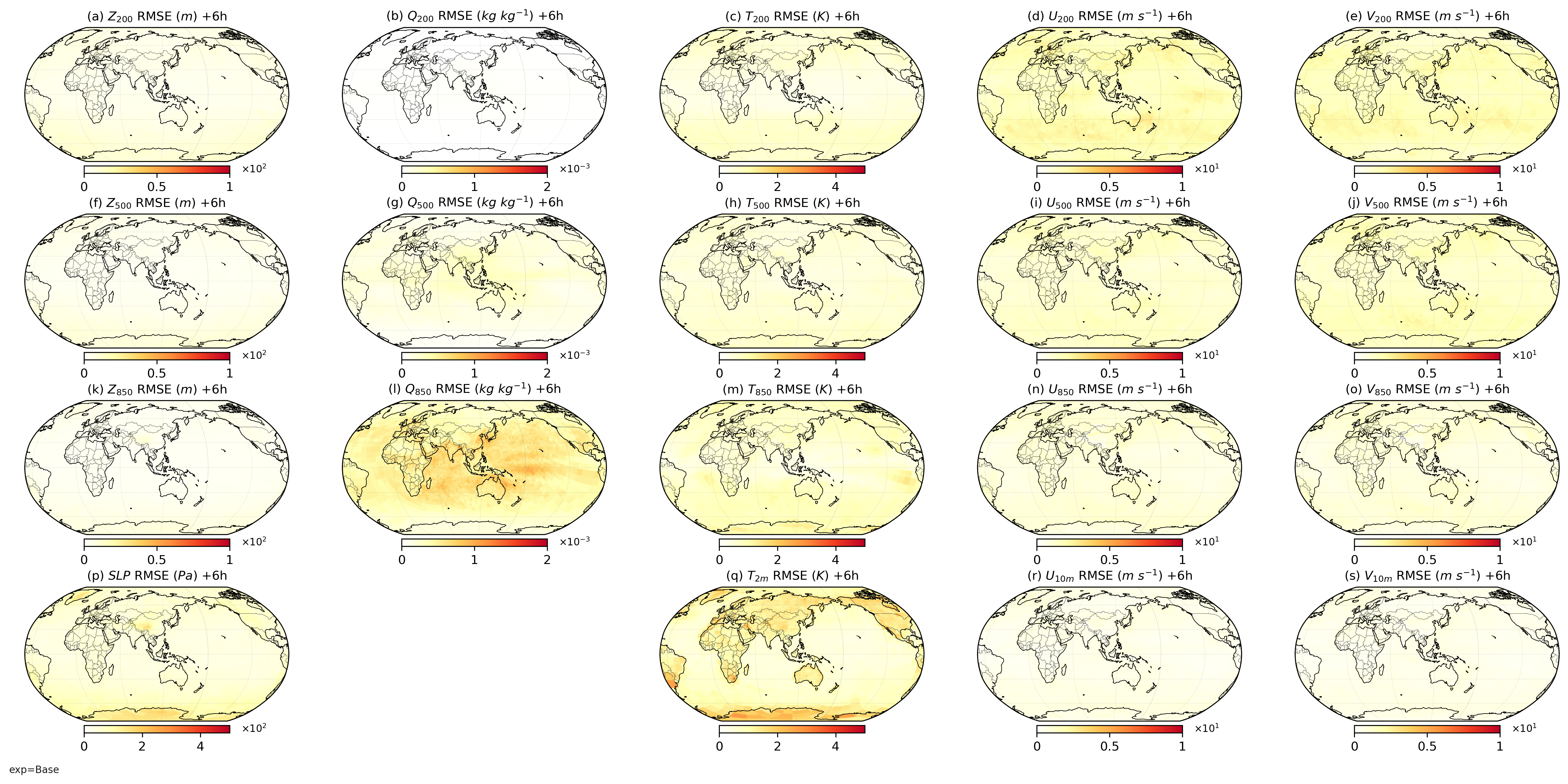}
\caption{Denormalized spatial \texttt{RMSE} distribution of the \texttt{SCS\_Base} one-step model over the test set at \texttt{+6 h}. Panels (a-e) show \texttt{Z200}, \texttt{Q200}, \texttt{T200}, \texttt{U200}, and \texttt{V200}; panels (f-j) show \texttt{Z500}, \texttt{Q500}, \texttt{T500}, \texttt{U500}, and \texttt{V500}; panels (k-o) show \texttt{Z850}, \texttt{Q850}, \texttt{T850}, \texttt{U850}, and \texttt{V850}; panel (p) shows \texttt{SLP}; panel (q) shows \texttt{T2m}; panel (r) shows \texttt{U10m}; and panel (s) shows \texttt{V10m}. The color bar below each panel gives the \texttt{RMSE} magnitude in the corresponding physical unit.}
\label{fig:fig4}
\end{figure}

\figlink{fig:fig4}{4} shows the spatial root mean squared error (\texttt{RMSE}) distribution of \texttt{SCS\_Base Model} at \texttt{+6 h}, i.e., in one-step prediction mode. The goal of this section is not to compare which variables are intrinsically easier or harder, but to examine whether the spatial error distribution still preserves a physically reasonable variable hierarchy and vertical structure. The result shows a clear hierarchy across both variables and pressure levels. Geopotential height (\texttt{Z}) is relatively stable at all three levels, with upper and lower levels often slightly larger than \texttt{500 hPa}, which is consistent with differences in absolute amplitude and synoptic variability across levels. Specific humidity (\texttt{Q}) shows the clearest vertical contrast, with error increasing strongly toward lower levels and following \texttt{Q850 > Q500 > Q200}. That behavior is expected because low-level humidity has the largest absolute magnitude, the strongest gradients, and the most active weather-related variability, so the same relative bias produces larger absolute \texttt{RMSE}. Temperature (\texttt{T}) is broadly similar across levels, but \texttt{200 hPa} and \texttt{850 hPa} are often slightly larger than \texttt{500 hPa}, reflecting stronger control from upper-level jets and lower-level surface and boundary-layer processes. Zonal wind (\texttt{U}) is usually more pronounced aloft, especially for \texttt{U200}, because upper-tropospheric westerly jets have larger wind speed and sharper horizontal gradients. Meridional wind (\texttt{V}) is smoother overall than \texttt{U}, but its upper level also tends to be slightly larger because planetary-wave and trough-ridge structures generate stronger north-south anomalies aloft.

\subsection{Ablation Experiments}\label{sec:results-base-ablation}
The structural ablation results of \texttt{SCS\_Base Model} are summarized in \tablink{tab:ablation-overall-6h-36h}{2} and \tablink{tab:ablation-keyvars-center}{3}, and \supptablink{tab:s1}{S1} provides the corresponding global key-variable results. \tablink{tab:ablation-overall-6h-36h}{2} already gives the most direct conclusion: the full \texttt{SCS\_Base Model} achieves the best mean \texttt{RMSE} and \texttt{ACC} both over the center-refined face and under global evaluation. At \texttt{36 h}, for example, global \texttt{RMSE} increases from \texttt{0.189} for the full model to \texttt{0.201} when using a \texttt{Swin transformer-based FaceMixer}, to \texttt{0.209} when removing \texttt{GlobalMixer}, and to \texttt{0.212} when removing \texttt{FaceMixer}, while the ranking by \texttt{ACC} remains consistent. One useful detail is that the \texttt{Swin Transformer-based FaceMixer} variant always lies between the full model and the two direct module-removal variants. That means it can recover part of the local mixing capacity, but still does not outperform the \texttt{MogaNet-based Face Mixer}. \tablink{tab:ablation-keyvars-center}{3} extends the same conclusion to key variables over the center-refined face. At both \texttt{6 h} and \texttt{36 h}, the full \texttt{SCS\_Base Model} remains best on \texttt{Q850}, \texttt{T2m}, \texttt{T850}, \texttt{U10m}, and \texttt{Z500}, while the three degraded variants all lose forecasting skill to different degrees. The global six-face results in \supptablink{tab:s1}{S1} show the same pattern.

The roles of \texttt{Face Mixer} and \texttt{Global Mixer} can be understood more physically. When \texttt{Global Mixer} is removed, the most obvious degradations at \texttt{36 h} occur in \texttt{T2m} and \texttt{U10m}, whose \texttt{RMSE} over the center-refined face increases by \texttt{38.4\%} and \texttt{22.5\%}, respectively. The same sensitivity is also visible in the global supplementary results. This indicates that near-surface temperature and wind do not depend only on local gradients; they still rely on large-scale thermodynamic background and circulation input from outside the center-refined face, so forecasting over the refined region cannot depend on intra-face information alone. By contrast, removing \texttt{FaceMixer} causes more widespread degradation and affects \texttt{Z500} and humidity-related metrics especially strongly. This shows that, even after cross-face background information enters the center region, the model still needs effective intra-face mixing to maintain smooth gradients, wave structure, and local detail on the stretched mesh. In that sense, \texttt{Global Mixer} prevents regional forecasting from becoming disconnected from the surrounding large-scale background, while \texttt{Face Mixer} organizes that background into useful regional temperature gradients, wind structure, and moist-region organization. Both are necessary.


\newcommand{\best}[1]{\textcolor{blue}{\textbf{#1}}}

\begin{table*}[t]
\centering
\setlength{\tabcolsep}{2.2pt}
\begin{minipage}[t]{0.48\textwidth}
\centering
\scriptsize
\begin{tabular}{lcccc}
\toprule
& Base & \makecell{Base w/\\SwinT\\FaceMixer} & \makecell{Base w/o\\FaceMixer} & \makecell{Base w/o\\GlobalMixer} \\
\midrule
Center RMSE $\downarrow$ 6h  & \best{0.115} & 0.123 & 0.127 & 0.125 \\
Center RMSE $\downarrow$ 36h & \best{0.210} & 0.222 & 0.231 & 0.230 \\
Global RMSE $\downarrow$ 6h  & \best{0.095} & 0.102 & 0.106 & 0.104 \\
Global RMSE $\downarrow$ 36h & \best{0.189} & 0.201 & 0.212 & 0.209 \\
\bottomrule
\end{tabular}
\end{minipage}
\hfill
\begin{minipage}[t]{0.48\textwidth}
\centering
\scriptsize
\begin{tabular}{lcccc}
\toprule
& Base & \makecell{Base w/\\SwinT\\FaceMixer} & \makecell{Base w/o\\FaceMixer} & \makecell{Base w/o\\GlobalMixer} \\
\midrule
Center ACC $\uparrow$ 6h  & \best{0.963} & 0.960 & 0.958 & 0.959 \\
Center ACC $\uparrow$ 36h & \best{0.930} & 0.924 & 0.918 & 0.918 \\
Global ACC $\uparrow$ 6h  & \best{0.964} & 0.962 & 0.961 & 0.961 \\
Global ACC $\uparrow$ 36h & \best{0.929} & 0.923 & 0.916 & 0.918 \\
\bottomrule
\end{tabular}
\end{minipage}
\caption{Overall ablation results at \texttt{6 h} and \texttt{36 h}. The full \texttt{SCS\_Base} model achieves the best mean \texttt{RMSE} and \texttt{ACC} both over the center-refined face and globally, indicating that both the \texttt{FaceMixer} and the \texttt{GlobalMixer} make stable positive contributions to forecast quality.}
\label{tab:ablation-overall-6h-36h}
\end{table*}

\begin{table*}[t]
\centering
\scriptsize
\setlength{\tabcolsep}{3.2pt}
\resizebox{\textwidth}{!}{%
\begin{tabular}{lcccccccc}
\toprule
\makecell[l]{Variable @ Lead} & \makecell{Base\\RMSE $\downarrow$} & \makecell{Base\\ACC $\uparrow$} & \makecell{Base w/ SwinT\\FaceMixer\\RMSE $\downarrow$} & \makecell{Base w/ SwinT\\FaceMixer\\ACC $\uparrow$} & \makecell{Base w/o\\FaceMixer\\RMSE $\downarrow$} & \makecell{Base w/o\\FaceMixer\\ACC $\uparrow$} & \makecell{Base w/o\\GlobalMixer\\RMSE $\downarrow$} & \makecell{Base w/o\\GlobalMixer\\ACC $\uparrow$} \\
\midrule
Q850 [kg kg$^{-1}$] @ 6h  & \best{0.001} & \best{0.985} & 0.001 (+6.0\%) & 0.983 (-0.2\%) & 0.001 (+9.5\%) & 0.982 (-0.3\%) & 0.001 (+8.0\%) & 0.982 (-0.3\%) \\
Q850 [kg kg$^{-1}$] @ 36h & \best{0.001} & \best{0.955} & 0.001 (+6.1\%) & 0.951 (-0.5\%) & 0.001 (+10.4\%) & 0.946 (-1.0\%) & 0.001 (+10.4\%) & 0.946 (-1.0\%) \\
T2m [K] @ 6h             & \best{1.113} & \best{0.994} & 1.233 (+10.7\%) & 0.992 (-0.1\%) & 1.269 (+14.0\%) & 0.992 (-0.2\%) & 1.329 (+19.3\%) & 0.991 (-0.3\%) \\
T2m [K] @ 36h            & \best{1.578} & \best{0.988} & 1.720 (+9.0\%) & 0.986 (-0.2\%) & 1.798 (+13.9\%) & 0.984 (-0.4\%) & 2.184 (+38.4\%) & 0.977 (-1.1\%) \\
T850 [K] @ 6h            & \best{0.666} & \best{0.997} & 0.724 (+8.7\%) & 0.996 (-0.1\%) & 0.750 (+12.6\%) & 0.996 (-0.1\%) & 0.756 (+13.4\%) & 0.996 (-0.1\%) \\
T850 [K] @ 36h           & \best{1.062} & \best{0.992} & 1.145 (+7.8\%) & 0.991 (-0.1\%) & 1.220 (+14.9\%) & 0.990 (-0.2\%) & 1.278 (+20.4\%) & 0.989 (-0.3\%) \\
U10 [m s$^{-1}$] @ 6h    & \best{0.679} & \best{0.969} & 0.739 (+8.9\%) & 0.963 (-0.6\%) & 0.769 (+13.3\%) & 0.960 (-0.9\%) & 0.782 (+15.2\%) & 0.958 (-1.1\%) \\
U10 [m s$^{-1}$] @ 36h   & \best{1.141} & \best{0.912} & 1.233 (+8.1\%) & 0.896 (-1.7\%) & 1.301 (+14.0\%) & 0.885 (-3.0\%) & 1.397 (+22.5\%) & 0.869 (-4.7\%) \\
Z500 [m] @ 6h            & \best{4.001} & \best{0.999} & 4.581 (+14.5\%) & 0.999 (-0.0\%) & 4.797 (+19.9\%) & 0.999 (-0.0\%) & 4.708 (+17.7\%) & 0.999 (-0.0\%) \\
Z500 [m] @ 36h           & \best{9.284} & \best{0.996} & 10.937 (+17.8\%) & 0.995 (-0.1\%) & 12.308 (+32.6\%) & 0.994 (-0.3\%) & 11.899 (+28.2\%) & 0.994 (-0.2\%) \\
\bottomrule
\end{tabular}%
}
\caption{Ablation results for key variables on the center-refined face at \texttt{6 h} and \texttt{36 h}. Consistent with the global key-variable results reported in Supplementary Table~S1, the full \texttt{SCS\_Base} model remains the best-performing variant over the center region.}
\label{tab:ablation-keyvars-center}
\end{table*}

\subsection{Case study}\label{sec:results-base-case}
\begin{figure}[H]
\centering
\includegraphics[width=\textwidth]{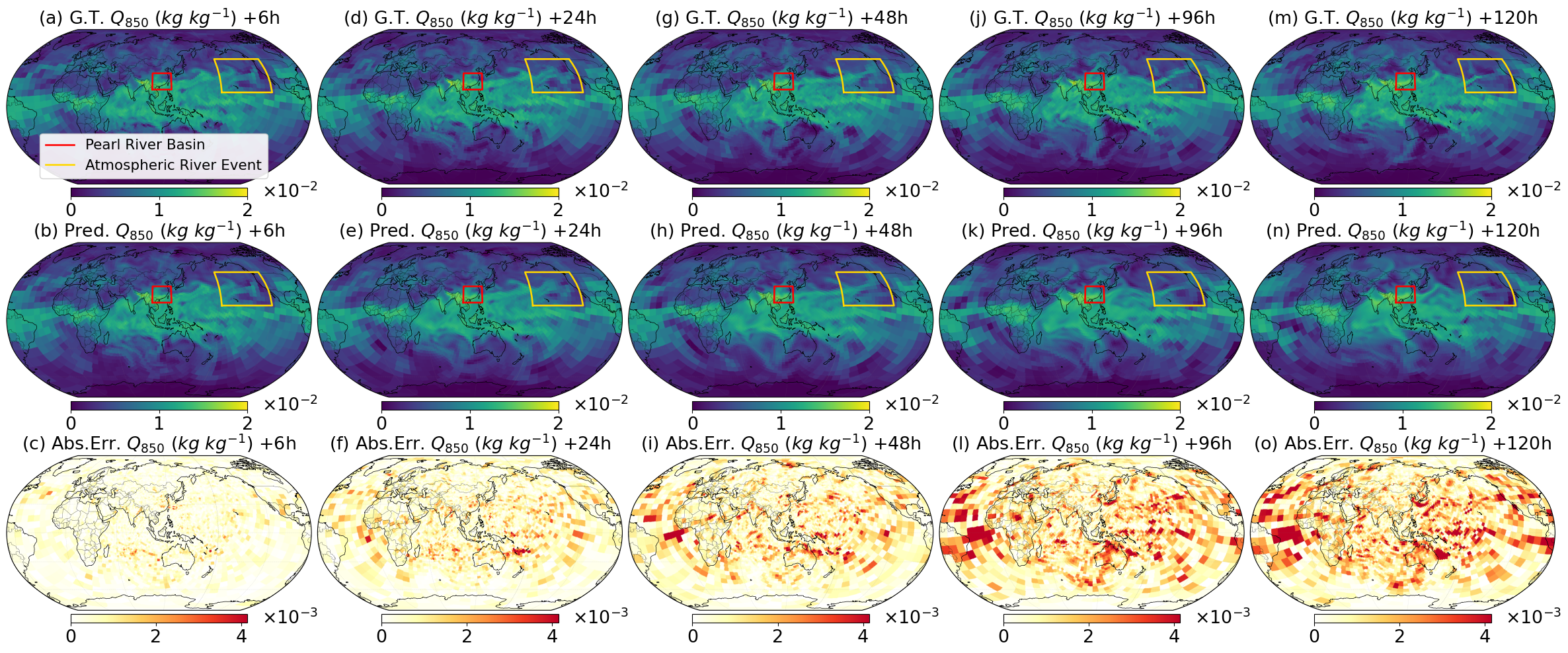}
\caption{Forecast of \texttt{850 hPa} specific humidity (\texttt{Q850}) for a representative case from the \texttt{SCS\_Base} one-step model. Panels (a-e) show the ground truth (\texttt{GT}) at \texttt{+6 h}, \texttt{+24 h}, \texttt{+48 h}, \texttt{+96 h}, and \texttt{+120 h}; panels (f-j) show the prediction (\texttt{Pred}) at the same lead times; and panels (k-o) show the absolute error (\texttt{Abs. Err.}) at the same lead times. The first two rows share the same field color scale, the third row uses an absolute-error color scale, and the text in the lower-left corner reports the experiment name and sample index. If highlighted boxes are present, they indicate two additionally annotated local weather-process regions during this period: the high-moisture region over the Pearl River Basin and the atmospheric river (\texttt{AR}) event extending from the eastern Pacific to North America.}
\label{fig:fig5}
\end{figure}

\texttt{SCS\_Base Model} is trained only for one-step prediction, so the long-range rollout shown here is not meant to treat it as a final medium-range forecasting system. Instead, it is used to test whether detail over the center-refined region collapses quickly under repeated forward integration on this cross-scale variable-resolution mesh. To answer that question, we choose \texttt{850 hPa} specific humidity (\texttt{Q850}) as the main diagnostic variable. \texttt{Q850} is closely tied to low-level moisture transport and precipitation processes, and its spatial texture is usually richer than that of \texttt{Z500} or \texttt{T850}, making it better suited to testing whether the model can still preserve weather-scale detail. \figlink{fig:fig5}{5} gives the forecast, ground truth, and absolute error for a representative case initialized at \texttt{2022-06-07 12:00}, including both the one-step \texttt{+6 h} result and longer lead times obtained by simple autoregressive rollout, so that it can later be compared directly with \texttt{SCS\_FCST4 Model}.

This case shows that detail over the central region does not collapse immediately during rollout. Because the framework uses coarser resolution outside the target region and finer resolution over the center-refined face, repeated cross-scale interaction could in principle smooth out local structure quickly. \figlink{fig:fig5}{5} shows, however, that even with the one-step model, the main moist region near South China is still identifiable after rollout to \texttt{5} days, and its spatial organization remains broadly continuous with the one-step \texttt{+6 h} result. At least for the present case and the present resolution, key information over the center-refined region does not immediately degrade into an overly smooth field as lead time increases.

That does not mean that error does not grow. Error still accumulates with lead time, but \figlink{fig:fig5}{5} also shows that this growth is spatially heterogeneous. Error expands more rapidly in the outer coarse-resolution regions, whereas the center-refined face, especially over eastern China, remains comparatively controlled. This suggests that center refinement is not merely a formal change in resolution. It genuinely helps the model maintain local structure over the target region for longer. That advantage, however, is reflected more strongly in the retention of regional detail than in equally strong skill for all long-distance transport processes.

Two weather processes in the same case make this point clearer. The first is the active \texttt{Dragon Boat Rain} period over the Pearl River Basin in \texttt{2022}. Previous work has shown that precipitation over this region was significantly above normal during \texttt{2022-05-21} to \texttt{2022-06-21} \citep{luo2025dragonboat}. Against that background, \figlink{fig:fig5}{5} shows that after initialization at \texttt{2022-06-07 12:00}, strong low-level moisture signals persist across multiple lead times over South China, especially northern Guangdong, northern Guangxi, and the northern Pearl River Basin. This indicates that \texttt{SCS\_Base Model} can already retain the main moisture structure associated with that event.

The second process is a strong atmospheric-river event near the eastern North Pacific during \texttt{2022-06-09} to \texttt{2022-06-12}, summarized by \citet{cw3e_eventsummary_june2022}. For such a long-distance transport process, \texttt{SCS\_Base Model} can already produce a clear, narrow, and continuous moisture band extending from the northeastern Pacific to the west coast of North America within \texttt{5} days. By around \texttt{2022-06-12}, however, that plume no longer extends toward the northwestern United States and southwestern Canada as clearly as in the ground truth. This result is already enough to show that cross-scale variable-resolution forecasting on this mesh is feasible and can retain a substantial amount of detail over the center region. But for processes that depend on sustained long-range transport and whose medium-range evolution is more path-sensitive, the capability of the one-step model remains limited. That limitation is one of the direct motivations for introducing rollout-oriented training.

\suppfiglink{fig:s3}{S3} and \suppfiglink{fig:s4}{S4} use the same initialization time and lead times as \figlink{fig:fig5}{5} to show \texttt{Z500} and \texttt{WS10m}, respectively. Viewed together with \figlink{fig:fig5}{5}, they indicate that \texttt{SCS\_Base Model} already provides a sufficiently stable representational foundation across low-level moisture, midlevel circulation, and near-surface wind, and is therefore worth developing further into \texttt{SCS\_FCST4 Model}.

\section{Results in SCS\_FCST4}\label{sec:results-fcst4}

\subsection{Averaged performances}\label{sec:results-fcst4-avg}
\texttt{SCS\_FCST4 Model} shifts the training objective from one-step error to consistency over a short forecast trajectory. A one-step model directly minimizes error only at the nearest future time, whereas errors at longer lead times are exposed only gradually through autoregressive rollout. By contrast, \texttt{SCS\_FCST4 Model} is penalized on four future times simultaneously in a shared latent space, so it must learn a state representation that remains useful across a longer forecast interval. That matters for \texttt{5-15} day forecasting, because medium-range stability depends not only on whether the first step is accurate, but also on whether errors in early rollout are quickly amplified.

\begin{figure}[H]
\centering
\includegraphics[width=\textwidth]{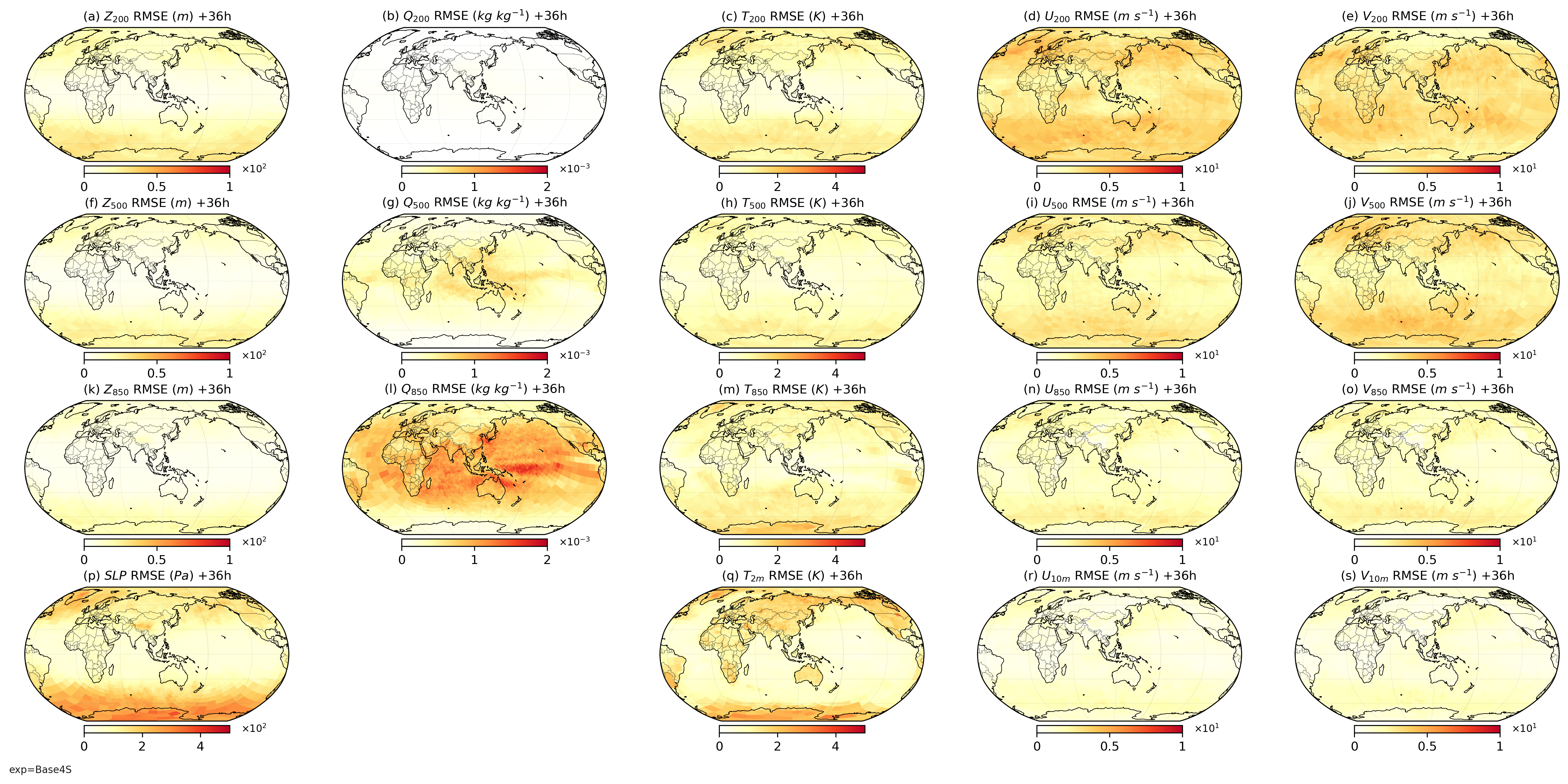}
\caption{Denormalized spatial \texttt{RMSE} distribution of the \texttt{SCS\_FCST4} model at \texttt{+36 h}. Panels (a-e) show \texttt{Z200}, \texttt{Q200}, \texttt{T200}, \texttt{U200}, and \texttt{V200}; panels (f-j) show \texttt{Z500}, \texttt{Q500}, \texttt{T500}, \texttt{U500}, and \texttt{V500}; panels (k-o) show \texttt{Z850}, \texttt{Q850}, \texttt{T850}, \texttt{U850}, and \texttt{V850}; panel (p) shows \texttt{SLP}; panel (q) shows \texttt{T2m}; panel (r) shows \texttt{U10m}; and panel (s) shows \texttt{V10m}. The color bar below each panel gives the \texttt{RMSE} magnitude in the corresponding physical unit.}
\label{fig:fig6}
\end{figure}

That change first appears in the spatial distribution of error. \figlink{fig:fig6}{6} shows the spatial \texttt{RMSE} distribution of \texttt{SCS\_FCST4 Model} at \texttt{+36 h}, while \suppfiglink{fig:s5}{S5} and \suppfiglink{fig:s6}{S6} provide the corresponding results at \texttt{+6 h} and \texttt{+120 h}. Viewed together, these figures show that error does accumulate with lead time, but the \texttt{RMSE} structure does not quickly degenerate into structureless noise. Different variables and levels still exhibit organized spatial patterns, and the main contrasts are still controlled by the magnitudes and spatial distributions of the variables themselves rather than by a single nonphysical diffusion mode.

\begin{figure}[H]
\centering
\includegraphics[width=\textwidth]{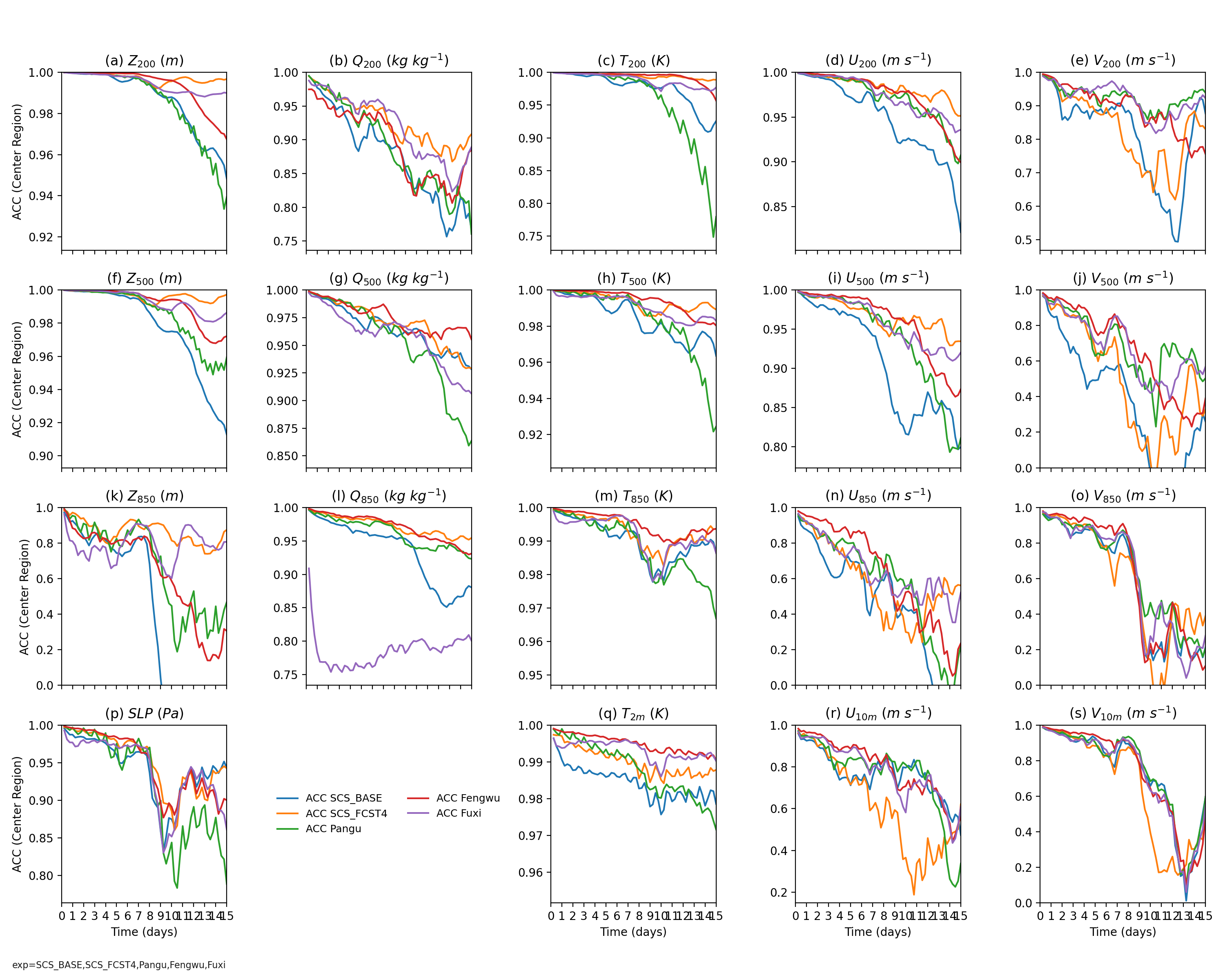}
\caption{Multi-model \texttt{ACC} time series over the center-refined face. Panels (a-e) show \texttt{Z200}, \texttt{Q200}, \texttt{T200}, \texttt{U200}, and \texttt{V200}; panels (f-j) show \texttt{Z500}, \texttt{Q500}, \texttt{T500}, \texttt{U500}, and \texttt{V500}; panels (k-o) show \texttt{Z850}, \texttt{Q850}, \texttt{T850}, \texttt{U850}, and \texttt{V850}; panel (p) shows \texttt{SLP}; panel (q) shows \texttt{T2m}; panel (r) shows \texttt{U10m}; and panel (s) shows \texttt{V10m}. The x-axis is time in days and the y-axis is \texttt{ACC} over the center-refined face. Blue, orange, green, red, and purple curves denote \texttt{SCS\_Base}, \texttt{SCS\_FCST4}, \texttt{Pangu}, \texttt{FengWu}, and \texttt{FuXi}, respectively, and the legend is placed in the blank panel on the bottom row.}
\label{fig:fig7}
\end{figure}

Spatial \texttt{RMSE} shows that the model does not become rapidly unstable at medium range, but cross-model comparison also requires a metric more robust to resolution mismatch. \figlink{fig:fig7}{7} therefore turns to the multi-model \texttt{ACC} evolution over the center-refined face. The comparison includes \texttt{SCS\_Base Model}, \texttt{SCS\_FCST4 Model}, \texttt{Pangu-Weather}, \texttt{FengWu}, and \texttt{FuXi}. Because these models differ in original resolution and mesh system, we use anomaly correlation coefficient (\texttt{ACC}) as the main comparison metric, since it is less directly amplified by resolution mismatch and resampling error than \texttt{RMSE}.

Across variable families, geopotential height and temperature have the highest \texttt{ACC} and decay most slowly, indicating that large-scale circulation and thermodynamic structure remain the most stable forecasting targets across models. For these variables, \texttt{SCS\_FCST4 Model} performs strongly on \texttt{Z200}, \texttt{Z500}, \texttt{Z850}, and \texttt{T500/T850}, with particularly clear improvement on \texttt{Z850}. At longer lead times, it maintains higher \texttt{ACC}, whereas \texttt{SCS\_Base Model} decays more quickly after about day \texttt{8}.

Humidity fields show the benefit of joint multistep training even more clearly. \texttt{SCS\_FCST4 Model} is among the stronger performers on \texttt{Q200} and \texttt{Q500}, and remains in the competitive group together with \texttt{FengWu} and \texttt{Pangu-Weather} on \texttt{Q850}. These results indicate that, after rollout-oriented training, the \texttt{SCS} route can preserve low- and midlevel moisture transport and the spatial organization of moist regions over the center-refined region competitively.

Wind remains the hardest variable family, especially the \texttt{V} component and lower-level winds, whose \texttt{ACC} generally declines faster after about day \texttt{8}. Even so, \texttt{SCS\_FCST4 Model} still outperforms \texttt{SCS\_Base Model} overall on \texttt{U200/U500}, and the gain on \texttt{U850} is also clear. By contrast, the behavior of \texttt{V200/V500/V850} and \texttt{V10m} depends more strongly on the specific variable. Among surface variables, \texttt{SLP} and \texttt{T2m} still retain relatively high \texttt{ACC}, and \texttt{SCS\_FCST4 Model} is consistently better than \texttt{SCS\_Base Model} on \texttt{T2m}.

\suppfiglink{fig:s7}{S7} extends the same comparison to all six global faces. Because the resolution outside the central region is much coarser, the global results should not be interpreted as a strict like-for-like competition against uniform global \texttt{0.25 degree} baselines. They still support the same conclusion, however: in the global setting, \texttt{SCS\_Base Model} is clearly weaker than the comparison models on almost all variables, whereas \texttt{SCS\_FCST4 Model} substantially narrows that gap and in some settings becomes comparable to or better than those heavier systems. The benefit of rollout-oriented multistep training therefore appears not only over the center-refined region, but also in global circulation consistency.

Taken together, \figlink{fig:fig7}{7} and \suppfiglink{fig:s7}{S7} support a simple conclusion. Even with a configuration that remains quite sparse, \texttt{SCS\_FCST4 Model} already maintains competitive \texttt{ACC} evolution over the center-refined region for large-scale geopotential height fields, specific humidity fields, and part of the lower-tropospheric winds, while also narrowing the gap to much heavier models at the global scale. Given that the model uses far fewer parameters, far less training compute, and substantially coarser resolution than those baselines, these results suggest that a center-refined variable-resolution spherical mesh does not substantially weaken regional medium-range predictability. Instead, it provides a viable framework for lightweight integrated global-regional forecasting.

\subsection{Case study}\label{sec:results-fcst4-case}
\begin{figure}[H]
\centering
\includegraphics[width=\textwidth]{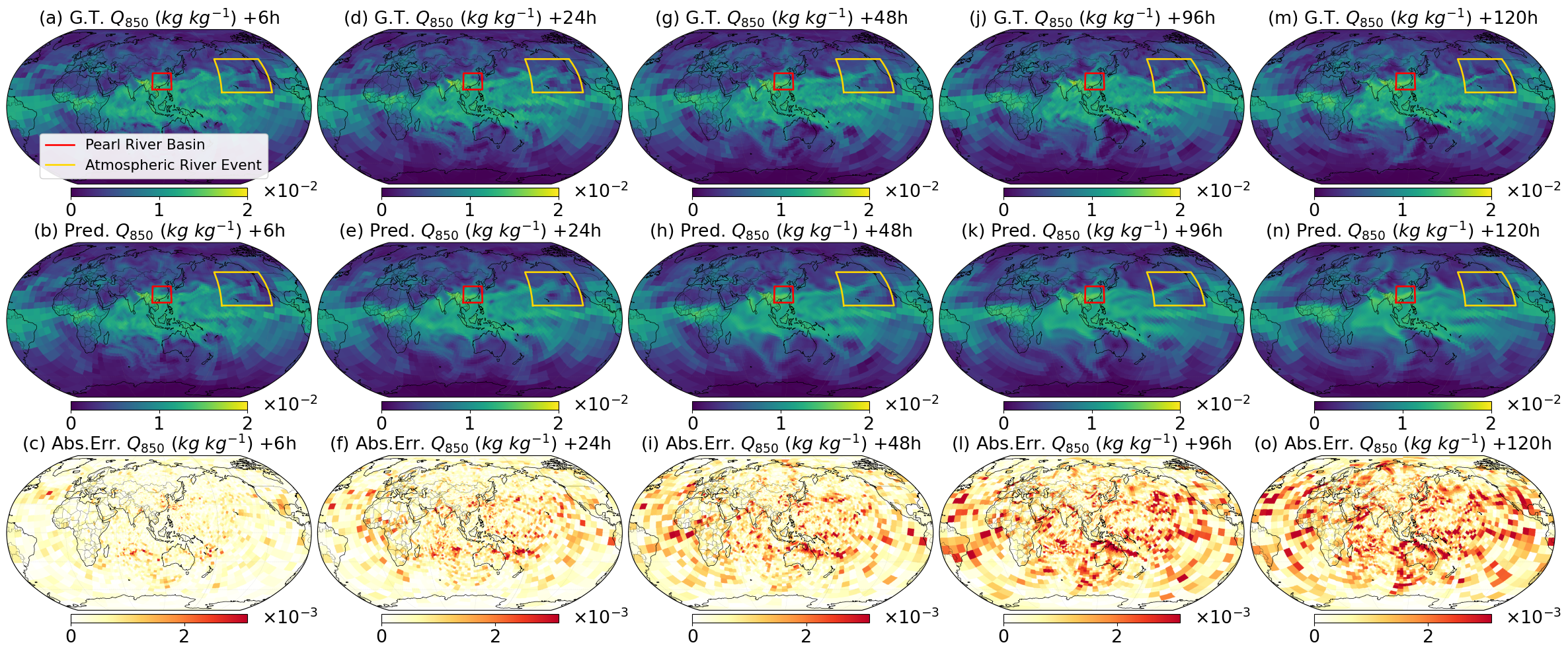}
\caption{Forecast of \texttt{Q850} for a representative case from the \texttt{SCS\_FCST4} model. Panels (a-e) show the ground truth (\texttt{GT}) at \texttt{+6 h}, \texttt{+24 h}, \texttt{+48 h}, \texttt{+96 h}, and \texttt{+120 h}; panels (f-j) show the prediction (\texttt{Pred}) at the same lead times; and panels (k-o) show the absolute error (\texttt{Abs. Err.}) at the same lead times. The first two rows share the same field color scale, the third row uses an absolute-error color scale. If highlighted boxes are present, they indicate additionally annotated local weather-process regions.}
\label{fig:fig8}
\end{figure}

\figlink{fig:fig8}{8} shows the \texttt{Q850} forecast from \texttt{SCS\_FCST4 Model} for the same case and the same lead times used in \figlink{fig:fig5}{5}, so the two figures provide a direct comparison. Relative to the one-step model, the advantage of \texttt{SCS\_FCST4 Model} lies mainly in preserving the temporal continuity of humidity structure rather than in coincidental closeness to a few local maxima. Over the boxed Pearl River Basin region in South China, the ground truth shows persistent and gradually organized low-level moist air across multiple lead times, corresponding to the background moisture conditions during the strong \texttt{Dragon Boat Rain} period of \texttt{2022}. Compared with the progressive loosening of structure and outward spread of error seen in \figlink{fig:fig5}{5} for the one-step model, \texttt{SCS\_FCST4 Model} keeps the position and extent of this moist core more stable. This suggests that, after joint multistep training, the model preserves low-level moisture anomalies tied to persistent heavy-rainfall conditions more effectively during rollout.

A similar improvement appears in the long, narrow moisture transport band extending from the northeastern Pacific toward the west coast of North America. \figlink{fig:fig5}{5} already showed that the one-step model can identify this structure at shorter lead times, but that its continuity and inland extent weaken noticeably at medium-range lead times. By contrast, \texttt{SCS\_FCST4 Model} still preserves the orientation, spatial continuity, and circulation-consistent organization of the band more completely after \texttt{+96 h}, making it closer to the atmospheric-river structure in the ground truth. More importantly, this improvement in structure is not achieved by merely shifting error elsewhere. It is accompanied by a reduction in overall error. Compared with \figlink{fig:fig5}{5}, the absolute error of \texttt{SCS\_FCST4 Model} is clearly smaller in amplitude, with the color-scale upper bound reduced from \texttt{4 x 10\^{}\{-3\}} to \texttt{2 x 10\^{}\{-3\}}. The difference is especially clear at \texttt{+96 h} and \texttt{+120 h}.

\suppfiglink{fig:s8}{S8} and \suppfiglink{fig:s9}{S9} further show that this gain is not limited to \texttt{Q850}. The error fields of \texttt{WS10m} and \texttt{Z500} are also clearly reduced relative to \texttt{SCS\_Base Model}, and the advantage is again more visible at longer lead times. The benefit of rollout-oriented training is therefore not confined to a single variable. It appears coherently across three complementary physical aspects: low-level humidity, near-surface wind, and midlevel circulation.

\subsection{Cross-face continuity under a real tropical cyclone}\label{sec:results-fcst4-typhoon}
In classical cubed-sphere and variable-resolution NWP studies, continuity across face boundaries is often tested with idealized boundary tests in which a \texttt{cosine bell}, \texttt{Gaussian hill}, or \texttt{slotted cylinder} crosses a panel edge or even a corner, so that numerical diffusion, phase error, and boundary artifacts can be examined directly \citep{putman2007finite,lauritzen2014transport}. Our object here is different. It is an AI forecasting model learned from real atmospheric evolution rather than a transport scheme that can be evaluated in isolation under an analytic tracer equation, so those idealized tests are not directly applicable. A typhoon case offers a more realistic continuity test instead. Typhoons contain a closed circulation, a near-circular wind core, and a clearly traceable moving path. Once such a system crosses a face boundary, discontinuity, displacement, or deformation becomes easy to identify visually.

\begin{figure}[H]
\centering
\includegraphics[width=\textwidth]{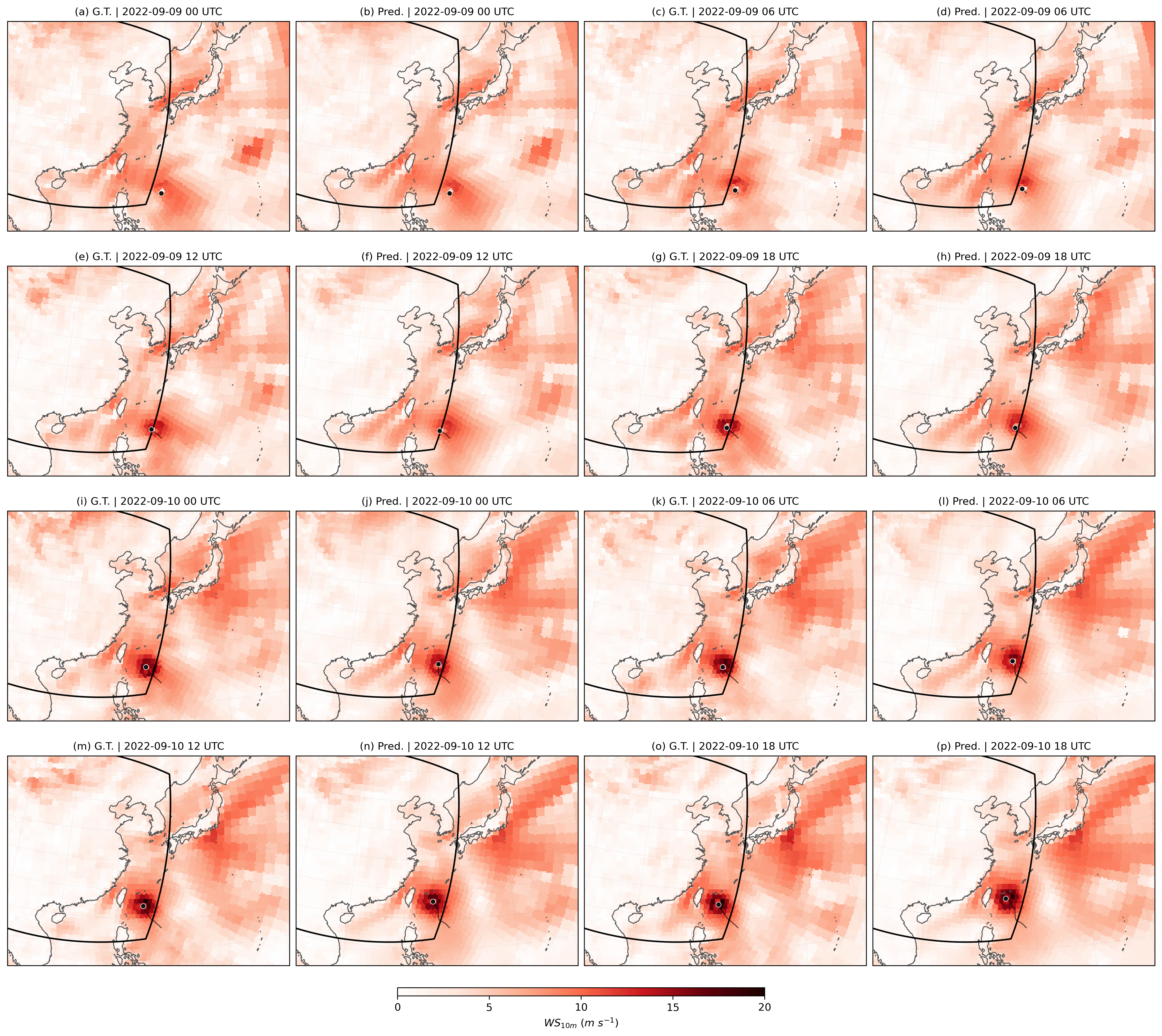}
\caption{Ground-truth and predicted \texttt{10 m} wind-speed fields (\texttt{WS10m}) for Typhoon Muifa from the \texttt{SCS\_FCST4} model. The \texttt{16} panels are arranged in chronological order, with each adjacent pair showing the ground truth (\texttt{GT}) and prediction (\texttt{Pred}) for the same valid time from \texttt{2022-09-09 00:00 UTC} to \texttt{2022-09-10 18:00 UTC} at \texttt{6 h} intervals. All panels share the same wind-speed color scale; thin black lines indicate the accumulated track, black dots mark the storm center at the current time, and black polygons show the outline of the center-refined face.}
\label{fig:fig9}
\end{figure}

\figlink{fig:fig9}{9} shows the case of Typhoon Muifa in September \texttt{2022}. The eastern boundary of the central region lies inside the active typhoon corridor of the northwestern Pacific, so Muifa provides a suitable test case for cross-face continuity. Its track and rapid-intensification phase can be checked against \citet{ibtracs2022muifa} and \citet{rammb2022muifa}. The sequence from \texttt{2022-09-09 00:00 UTC} to \texttt{2022-09-10 18:00 UTC} covers \texttt{8} successive \texttt{6 h} times during the key intensification stage and captures the storm as it moves northwestward from the active oceanic region east of the central face toward and across the boundary-adjacent area. The result shows that \texttt{SCS\_FCST4 Model} preserves geometric continuity across the boundary throughout the full period. The strong-wind core and the outer wind-speed gradient do not show visible fracture, displacement, or unnatural stitching as they approach and cross the face boundary, indicating that the transition between the central face and adjacent faces remains smooth for a real weather system.

The model also does more than capture the approximate storm outline. It reproduces the continued intensification of Muifa during this phase, with both the wind-radius extent and the inner-core strong-wind region growing clearly in time. Its northwestward track also remains highly consistent with the ground truth, with no obvious recursive drift or turning error. More importantly, although the mesh resolution here is not designed to resolve the finest typhoon-scale detail, the predicted field still preserves the asymmetric wind structure, including the relatively weaker wind-speed region on the western side of the storm and its contrast with the stronger wind band on the eastern side. This means the model captures not only the primary position and intensity of the storm, but also part of the secondary structure associated with environmental steering flow and circulation asymmetry. The significance of \figlink{fig:fig9}{9} is therefore not limited to presenting another successful typhoon case. It provides a physically relevant analogue to the classic boundary test: even for a high-gradient, fast-moving, boundary-crossing system such as a typhoon, \texttt{SCS\_FCST4 Model} can maintain cross-face continuity, a realistic storm track, and physically consistent intensity evolution at the same time.

\subsection{Spectral diagnostics}\label{sec:results-fcst4-psd}
Deterministic, purely data-driven weather models face a common risk in autoregressive forecasting: they can gradually lose small- and medium-scale perturbation energy, producing overly rapid spectral-tail decay and overly smooth forecast fields. Large-scale background flow may still remain, but the finer-scale structure tied to real weather processes can fade away. To test whether the model truly preserves multiscale dynamical and thermodynamic information, we therefore examine it from a spectral perspective.

\begin{figure}[H]
\centering
\includegraphics[width=\textwidth]{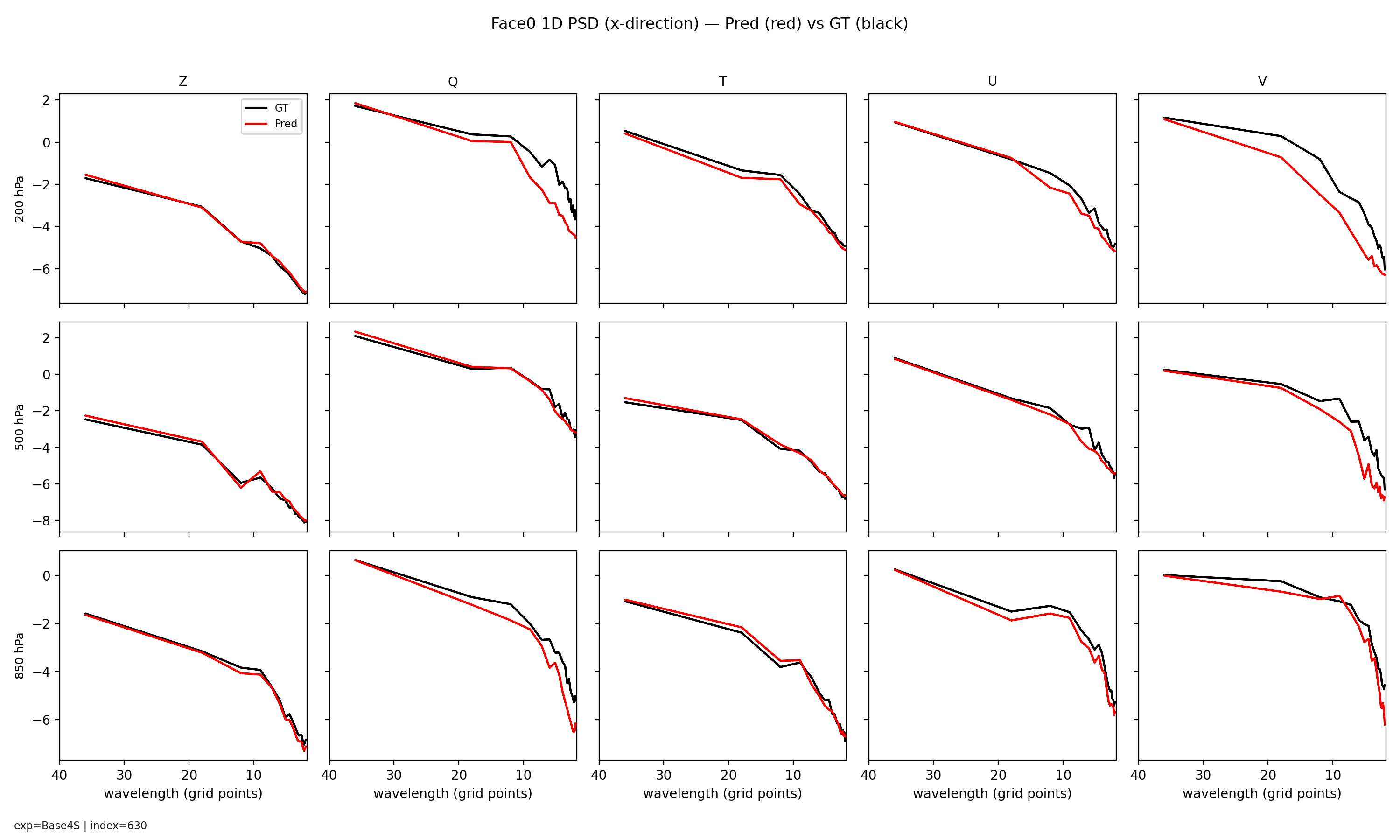}
\caption{Power spectral density (\texttt{PSD}) comparison on the center-refined face for the \texttt{SCS\_FCST4} model at \texttt{+36 h}. Columns correspond to \texttt{Z}, \texttt{Q}, \texttt{T}, \texttt{U}, and \texttt{V}, and rows correspond to \texttt{200 hPa}, \texttt{500 hPa}, and \texttt{850 hPa}. In each panel, the black curve denotes the ground truth (\texttt{GT}) and the red curve denotes the prediction (\texttt{Pred}); the x-axis is wavelength in grid points and the y-axis is log spectral energy.}
\label{fig:fig10}
\end{figure}

\figlink{fig:fig10}{10} shows the power spectral density (\texttt{PSD}) over the center-refined region at \texttt{+36 h}, covering \texttt{Z}, \texttt{Q}, \texttt{T}, \texttt{U}, and \texttt{V} at \texttt{200}, \texttt{500}, and \texttt{850 hPa}. The corresponding longer-lead results appear in \suppfiglink{fig:s10}{S10} at \texttt{+120 h} and \suppfiglink{fig:s11}{S11} at \texttt{+240 h}. Overall, the predicted spectra from \texttt{SCS\_FCST4 Model} remain close to the ground-truth spectra in both slope and energy level from large to medium scales, without showing the systematic shortwave collapse often seen in purely data-driven models. Geopotential height and temperature are the most stable variables. At both upper and lower levels, the predicted spectra of \texttt{Z} and \texttt{T} evolve almost parallel to the ground-truth spectra and remain closely aligned over most scales, with only limited underestimation at the shortest wavelengths. This indicates that the model preserves not only the large-scale circulation backbone but also much of the mesoscale variability associated with synoptic weather structure.

Humidity shows somewhat larger spectral departure than \texttt{Z} and \texttt{T}, especially at the shortwave end of \texttt{850 hPa}, where energy deficit is more visible. That is consistent with the stronger intermittency and sharper local gradients of the moisture field itself. Even so, the overall spectral shape and break structure remain close to the ground truth, indicating that the model does not simply flatten humidity into a single background field. Wind is the most challenging variable family. Both \texttt{U} and \texttt{V} show larger shortwave departure, especially aloft, reflecting the fact that small-scale momentum perturbations remain among the hardest components for deterministic data-driven models to reconstruct fully. Even in that case, however, the predicted spectra preserve the correct scale hierarchy and dominant slope, and do not collapse into a nonphysical distortion that worsens abruptly toward smaller scales.

The most important conclusion supported by \figlink{fig:fig10}{10} is therefore not that the model matches the ground truth perfectly at every variable and every wavelength. A more relevant result is that, as a purely data-driven predictor, \texttt{SCS\_FCST4 Model} does not show obvious small- and medium-scale energy dissipation at \texttt{36 h}, and on \texttt{Z} and \texttt{T} it already preserves the spectrum from synoptic to mesoscale structure quite stably. \suppfiglink{fig:s10}{S10} and \suppfiglink{fig:s11}{S11} further show that this is not confined to short lead times. As lead time increases from \texttt{36 h} to \texttt{120 h} and \texttt{240 h}, the spectral-tail bias does not show a monotonic, continuously collapsing trend. Instead, the predicted spectra of most variables remain within an error range similar to that already seen at \texttt{+36 h}. The model can therefore underestimate some shortwave energy in part of the variable space without allowing that bias to amplify uncontrollably in long rollout. In other words, rollout training in \texttt{SCS\_FCST4 Model} improves not only average error, but also the retention of at least part of the small- and medium-scale information.

\section{Conclusion and Limitation}\label{sec:conclusion}
The goal of this paper is not to continue scaling up heavier uniform global high-resolution models, but to test whether a global-regional representation with a center-refined face is already sufficient to support meaningful AI weather forecasting. From that perspective, the results of \texttt{StretchCast} can be summarized in two points. First, \texttt{SCS} is no longer just a geometrically plausible idea. \texttt{SCS\_Base Model} shows that, under a distinctly lightweight configuration with about \texttt{7,776} effective global grid cells and a resolution of about \texttt{0.875 degree} over the center-refined face, this representation already yields stable multivariate forecasts, and both \texttt{Face Mixer} and \texttt{Global Mixer} make measurable contributions to forecast quality. Second, \texttt{SCS\_FCST4 Model} shows that rollout-oriented multistep training can extend that representational advantage into medium-range lead times. The model maintains competitive \texttt{ACC} evolution over the center-refined region for large-scale geopotential height fields, specific humidity fields, and part of the lower-tropospheric winds, while also preserving cross-face continuity and multiscale structure in a real typhoon case and in spectral diagnostics. These results suggest that the \texttt{StretchCast} framework is worth pursuing further. Further increases in center resolution, training data volume, and temporal modeling strength may lead to better forecast performance. In addition, the simultaneous improvement of \texttt{SCS\_FCST4 Model} over the center-refined face and in the global setting suggests that the stretched cubed sphere offers a middle path between uniform global grids and limited-area forecasting. It preserves the large-scale circulation constraint of a closed global domain while allocating more computation to the target region. This is also consistent with the broader evolution of NWP models from fixed-resolution meshes toward variable-resolution designs.

This conclusion still needs to be interpreted within clear boundaries. The present \texttt{SCS} mesh remains quite sparse, so the paper demonstrates feasibility rather than the final resolution ceiling of the approach. The comparison with \texttt{Pangu-Weather}, \texttt{FengWu}, and \texttt{FuXi} is based on publicly available inference products after unified reprojection, which makes it more appropriate as a representation-level comparison than as a strict same-condition benchmark race. At the same time, the present work studies only one configuration whose center-refined face is centered over eastern China, and evaluates it mainly on \texttt{ERA5} reanalysis. The benefit of the same design over other target regions, its consistency with real observations, its probabilistic forecasting ability, and its stability under extreme events all still need further validation. That is why we treat the current results as a clear starting point rather than an endpoint. They show that this eastern-China-centered stretched cubed sphere, and more broadly the stretched-cubed-sphere route, is already worth developing further as a foundation for global-regional AI weather forecasting. The next step is to test how far it can be pushed under higher resolution, stronger models, and diffusion-based downscaling over the center region.

\subsection*{Acknowledgement}
This study is supported by the National Science Foundation of China (\texttt{42275009}).

\clearpage
\setcounter{figure}{0}
\setcounter{table}{0}
\renewcommand{\thefigure}{S\arabic{figure}}
\renewcommand{\thetable}{S\arabic{table}}

\section*{Supplementary Materials of ``StretchCast: Global-Regional AI Weather Forecasting on Stretched Cubed-Sphere Mesh''}

\subsection*{S1. Data mapping and geometric conditioning inputs}
\begin{figure}[H]
\centering
\includegraphics[width=\textwidth]{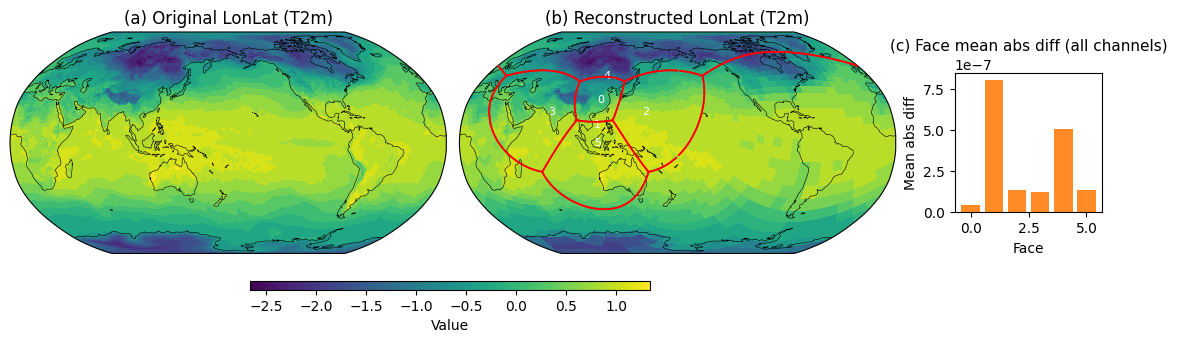}
\caption{Original \texttt{2 m} temperature (\texttt{T2m}) field and its reconstruction from the \texttt{SCS} representation. (a) Original latitude-longitude \texttt{T2m} field; (b) field reconstructed back from the \texttt{SCS} representation, with red curves marking face boundaries and white numbers denoting face indices; and (c) face-wise mean absolute error. The two map panels share the same color scale.}
\label{fig:s1}
\end{figure}

This section supplements \seclink{sec:data-methods}{3} of the main paper by addressing two basic issues related to the \texttt{SCS} representation. The first is whether the main spatial structure of a real meteorological field can still be preserved stably when it is mapped from a latitude-longitude grid to \texttt{SCS} and then reconstructed back again. \suppfiglink{fig:s1}{S1} addresses that question directly. It extends the geometric diagnosis in \figlink{fig:fig2}{2} of the main text, which used an idealized field, to a real \texttt{T2m} field, and further shows that the dominant source of mapping error is the resolution configuration itself rather than geometric fracture at face boundaries. Error is smaller over the center-refined region and larger over remote regions, which is consistent with the mesh design. More importantly, the reconstruction does not show obvious visual discontinuity across face boundaries, so the \texttt{SCS} representation can preserve the main structure of a real meteorological field in a stable way.

\begin{figure}[H]
\centering
\includegraphics[width=0.86\textwidth]{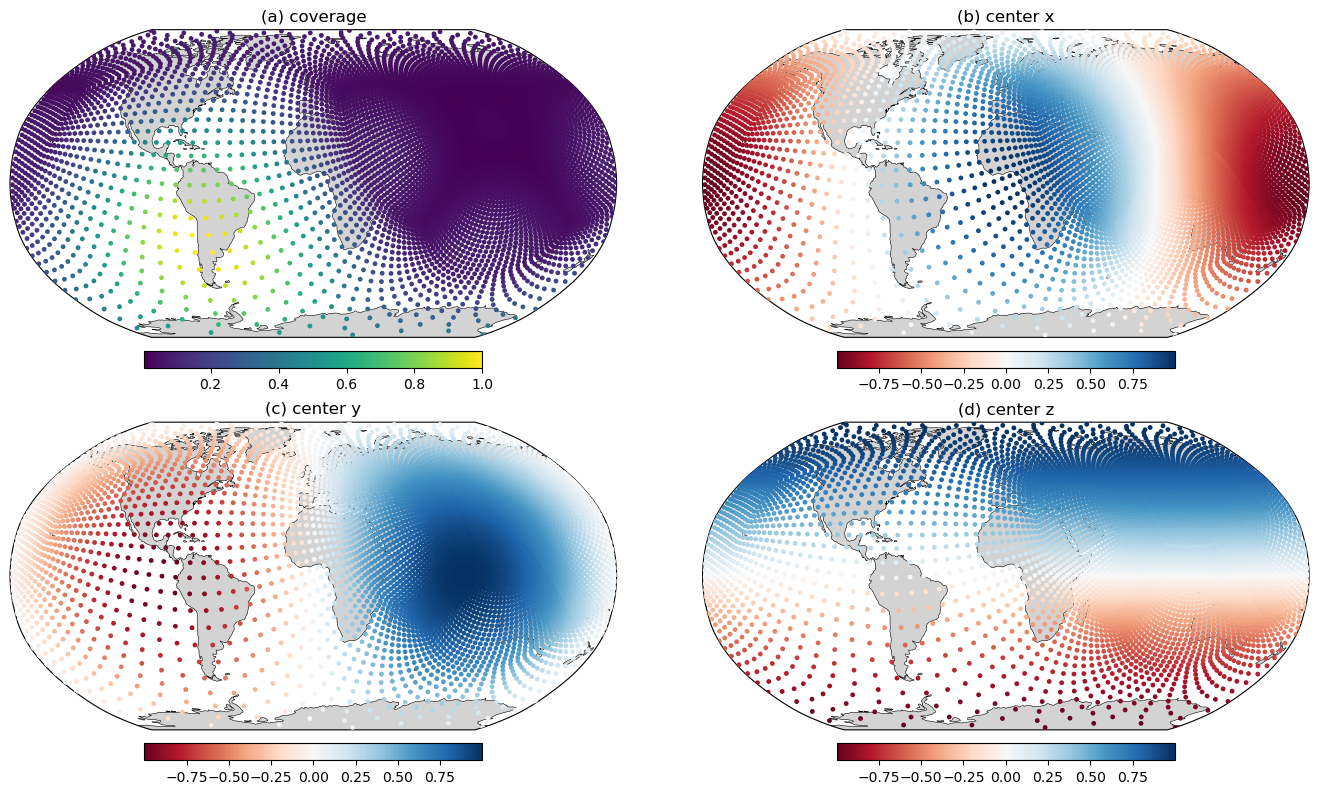}
\caption{Spherical-position input fields used for \texttt{FiLM} modulation. (a) \texttt{coverage}; (b) \texttt{center x}; (c) \texttt{center y}; and (d) \texttt{center z}. Each dot represents one \texttt{SCS} grid point over a global land-sea background. Colors in (a) show the \texttt{coverage} value, and colors in (b)-(d) show the three Cartesian components of the grid-center coordinates.}
\label{fig:s2}
\end{figure}

This geometric foundation matters because the later models do not operate on a uniform sphere. They operate on a strongly nonuniform spherical mesh while still using one shared tensor backbone. Shared token mixers and encoder-decoder blocks alone cannot automatically distinguish the geometric differences between the refined center region and the sparse outer region. \suppfiglink{fig:s2}{S2} therefore provides the geometric conditioning fields used by \texttt{3D Spherical Position FiLM}, including \texttt{coverage} and the 3D coordinates of the grid-center points on the unit sphere. The first describes local coverage and stretching, and the second encodes absolute spherical position as a continuous geometric signal. Because these conditions participate in layerwise modulation independently, the model can bring mesh nonuniformity directly into feature transformation rather than leaving it to be inferred implicitly by the network.

\subsection*{S2. Additional results for SCS\_Base}
This section supplements \seclink{sec:results-base}{4} of the main paper and expands the conclusions for \texttt{SCS\_Base Model} in two ways. The first concerns structural stability. \tablink{tab:ablation-overall-6h-36h}{2} and \tablink{tab:ablation-keyvars-center}{3} in the main text already show that the full \texttt{SCS\_Base} model is best on the central face and on key variables, while \supptablink{tab:s1}{S1} extends that conclusion to representative variables under global evaluation. The purpose of this table is not to repeat the result reported in the main text, but to show that the advantage of the full Base architecture does not depend on a single evaluation region. Whether the metric is computed over the region of interest or in a global average sense, removing \texttt{Face Mixer}, removing \texttt{Global Mixer}, or replacing \texttt{MogaNet} with a generic \texttt{SwinT FaceMixer} all leads to stable degradation.

\begin{figure}[H]
\centering
\includegraphics[width=\textwidth]{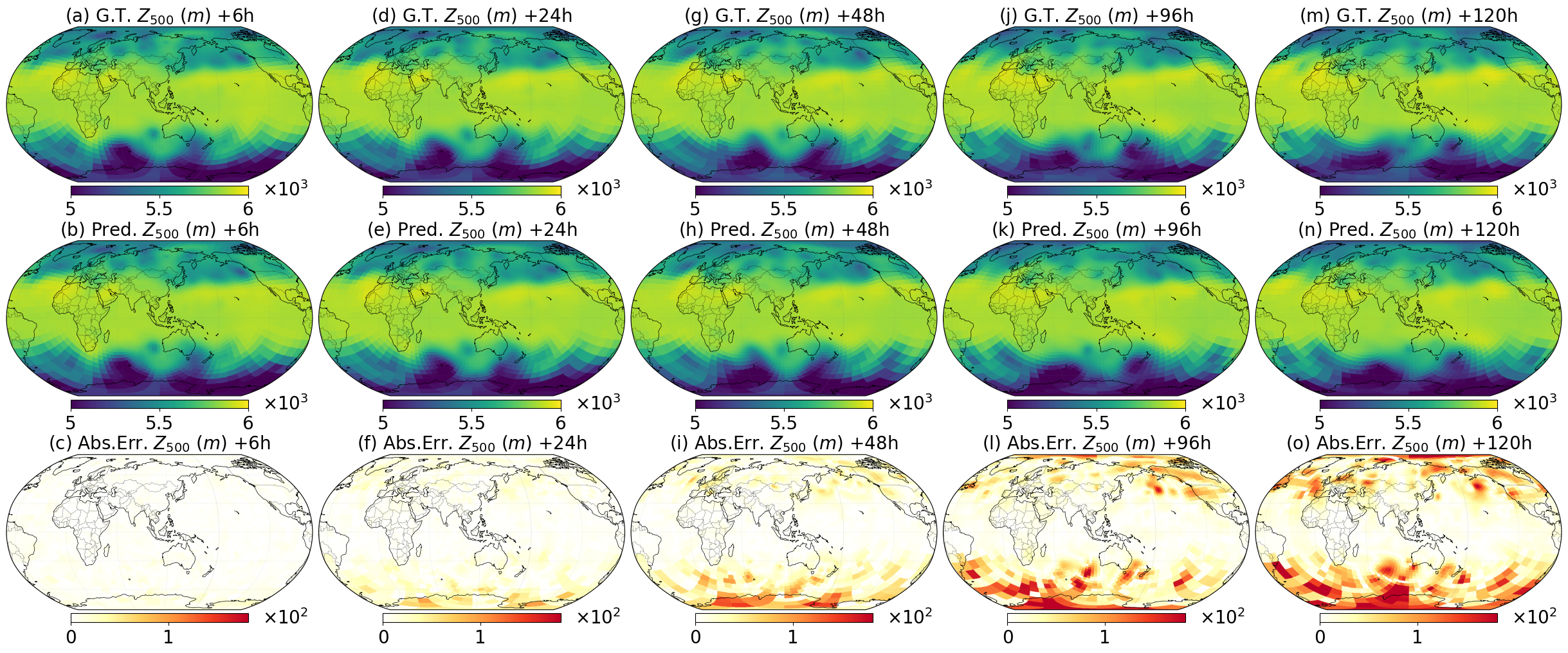}
\caption{Forecast of \texttt{500 hPa} geopotential height (\texttt{Z500}) from the \texttt{SCS\_Base} one-step model for the same representative case used in Fig.~5. Panels (a-e) show the ground truth (\texttt{GT}) at \texttt{+6 h}, \texttt{+24 h}, \texttt{+48 h}, \texttt{+96 h}, and \texttt{+120 h}; panels (f-j) show the prediction (\texttt{Pred}) at the same lead times; and panels (k-o) show the absolute error (\texttt{Abs. Err.}) at the same lead times. The first two rows share the same field color scale, the third row uses an absolute-error color scale, and the text in the lower-left corner reports the experiment name and sample index.}
\label{fig:s3}
\end{figure}

\begin{figure}[H]
\centering
\includegraphics[width=\textwidth]{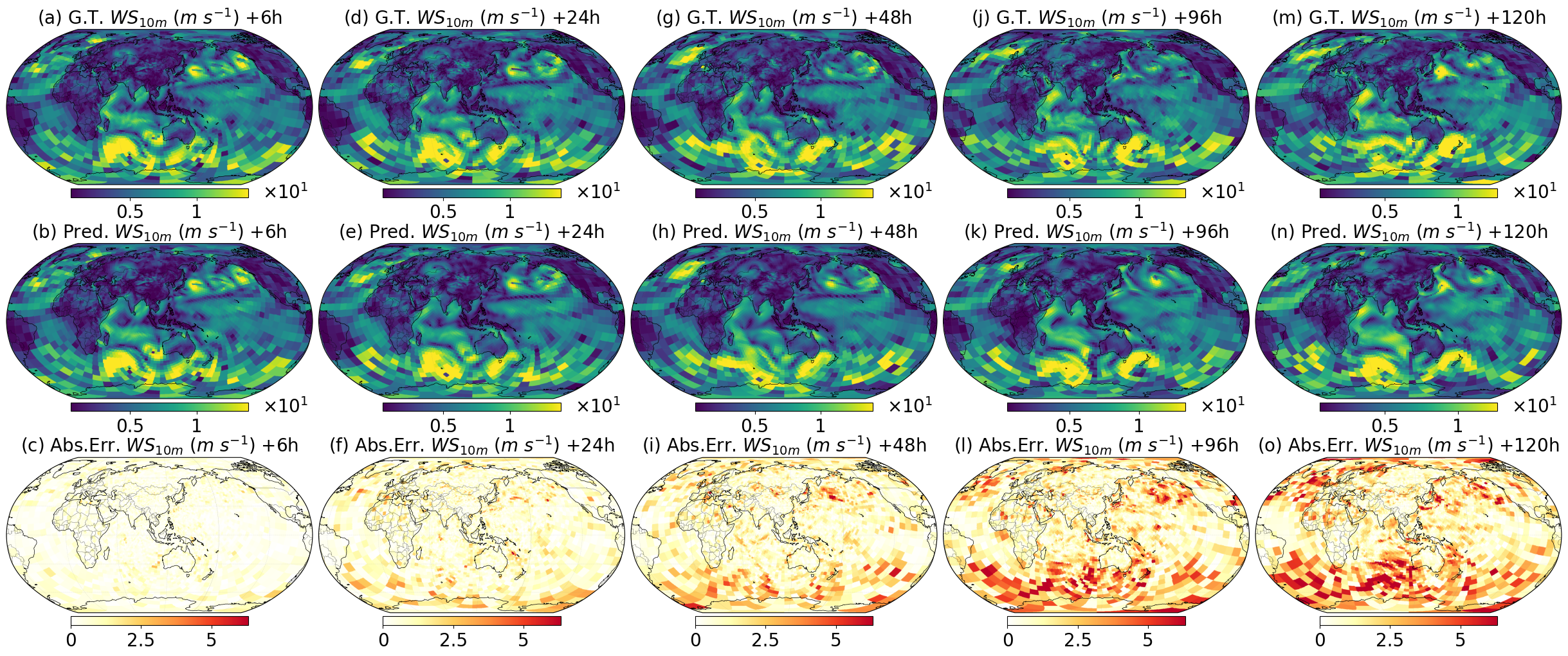}
\caption{Forecast of \texttt{10 m} wind speed (\texttt{WS10m}) from the \texttt{SCS\_Base} one-step model for the same representative case used in Fig.~5. Panels (a-e) show the ground truth (\texttt{GT}) at \texttt{+6 h}, \texttt{+24 h}, \texttt{+48 h}, \texttt{+96 h}, and \texttt{+120 h}; panels (f-j) show the prediction (\texttt{Pred}) at the same lead times; and panels (k-o) show the absolute error (\texttt{Abs. Err.}) at the same lead times. The first two rows share the same field color scale, the third row uses an absolute-error color scale, and the text in the lower-left corner reports the experiment name and sample index.}
\label{fig:s4}
\end{figure}

The second supplement concerns physical consistency at the individual-case level. \figlink{fig:fig5}{5} in the main text uses \texttt{Q850} to show that \texttt{SCS\_Base} can preserve low-level humidity structure in a representative case, but low-level humidity alone does not cover the full range of physical behavior. \suppfiglink{fig:s3}{S3} and \suppfiglink{fig:s4}{S4} therefore extend the same initialization time and the same lead times to \texttt{Z500} and \texttt{WS10m}. The parallel comparison shows more clearly that the feasibility of \texttt{SCS\_Base} is not confined to a single variable. It appears simultaneously in the midlevel circulation backbone, the near-surface wind field, and the low-level moisture structure. Its error still grows with lead time, but the main structures do not collapse quickly, which is exactly why the main paper then turns to \texttt{SCS\_FCST4}.

\phantomsection
\label{tab:s1}
\begin{table*}[H]
\centering
\scriptsize
\setlength{\tabcolsep}{3.2pt}
\resizebox{\textwidth}{!}{%
\begin{tabular}{lcccccccc}
\toprule
\makecell[l]{Variable @ Lead} & \makecell{Base\\RMSE $\downarrow$} & \makecell{Base\\ACC $\uparrow$} & \makecell{Base + SwinT\\FaceMixer\\RMSE $\downarrow$} & \makecell{Base + SwinT\\FaceMixer\\ACC $\uparrow$} & \makecell{Base w/o\\FaceMixer\\RMSE $\downarrow$} & \makecell{Base w/o\\FaceMixer\\ACC $\uparrow$} & \makecell{Base w/o\\GlobalMixer\\RMSE $\downarrow$} & \makecell{Base w/o\\GlobalMixer\\ACC $\uparrow$} \\
\midrule
Q850 [kg kg$^{-1}$] @ 6 h  & \best{0.001} & \best{0.979} & 0.001 (+6.2\%) & 0.976 (-0.3\%) & 0.001 (+10.2\%) & 0.974 (-0.5\%) & 0.001 (+9.0\%) & 0.975 (-0.4\%) \\
Q850 [kg kg$^{-1}$] @ 36 h & \best{0.001} & \best{0.925} & 0.001 (+5.7\%) & 0.917 (-0.9\%) & 0.001 (+10.9\%) & 0.908 (-1.9\%) & 0.001 (+10.3\%) & 0.909 (-1.7\%) \\
T2m [K] @ 6 h             & \best{0.988} & \best{0.993} & 1.096 (+10.9\%) & 0.991 (-0.2\%) & 1.139 (+15.3\%) & 0.991 (-0.2\%) & 1.204 (+21.9\%) & 0.990 (-0.3\%) \\
T2m [K] @ 36 h            & \best{1.490} & \best{0.984} & 1.616 (+8.5\%) & 0.982 (-0.3\%) & 1.696 (+13.8\%) & 0.980 (-0.5\%) & 2.017 (+35.4\%) & 0.972 (-1.3\%) \\
T850 [K] @ 6 h            & \best{0.639} & \best{0.996} & 0.698 (+9.3\%) & 0.995 (-0.1\%) & 0.728 (+14.1\%) & 0.995 (-0.1\%) & 0.724 (+13.3\%) & 0.995 (-0.1\%) \\
T850 [K] @ 36 h           & \best{1.102} & \best{0.988} & 1.190 (+7.9\%) & 0.986 (-0.2\%) & 1.274 (+15.6\%) & 0.984 (-0.4\%) & 1.293 (+17.3\%) & 0.984 (-0.5\%) \\
U10 [m s$^{-1}$] @ 6 h    & \best{0.639} & \best{0.984} & 0.691 (+8.3\%) & 0.981 (-0.3\%) & 0.721 (+12.8\%) & 0.979 (-0.5\%) & 0.729 (+14.2\%) & 0.979 (-0.5\%) \\
U10 [m s$^{-1}$] @ 36 h   & \best{1.277} & \best{0.936} & 1.386 (+8.5\%) & 0.924 (-1.2\%) & 1.475 (+15.5\%) & 0.914 (-2.3\%) & 1.522 (+19.2\%) & 0.909 (-2.8\%) \\
Z500 [m] @ 6 h            & \best{4.353} & \best{0.999} & 4.961 (+14.0\%) & 0.999 (-0.0\%) & 5.253 (+20.7\%) & 0.999 (-0.0\%) & 5.182 (+19.0\%) & 0.999 (-0.0\%) \\
Z500 [m] @ 36 h           & \best{13.229} & \best{0.994} & 15.175 (+14.7\%) & 0.992 (-0.2\%) & 16.574 (+25.3\%) & 0.991 (-0.3\%) & 15.944 (+20.5\%) & 0.992 (-0.3\%) \\
\bottomrule
\end{tabular}%
}
\caption{Global key-variable ablation results at \texttt{6 h} and \texttt{36 h}. This supplementary table shows that the full \texttt{SCS\_Base} model remains the best-performing variant on the same representative variables under global evaluation.}
\label{tab:ablation-keyvars-global-supp}
\end{table*}

\subsection*{S3. Additional results for averaged performance of SCS\_FCST4}
\begin{figure}[H]
\centering
\includegraphics[width=\textwidth]{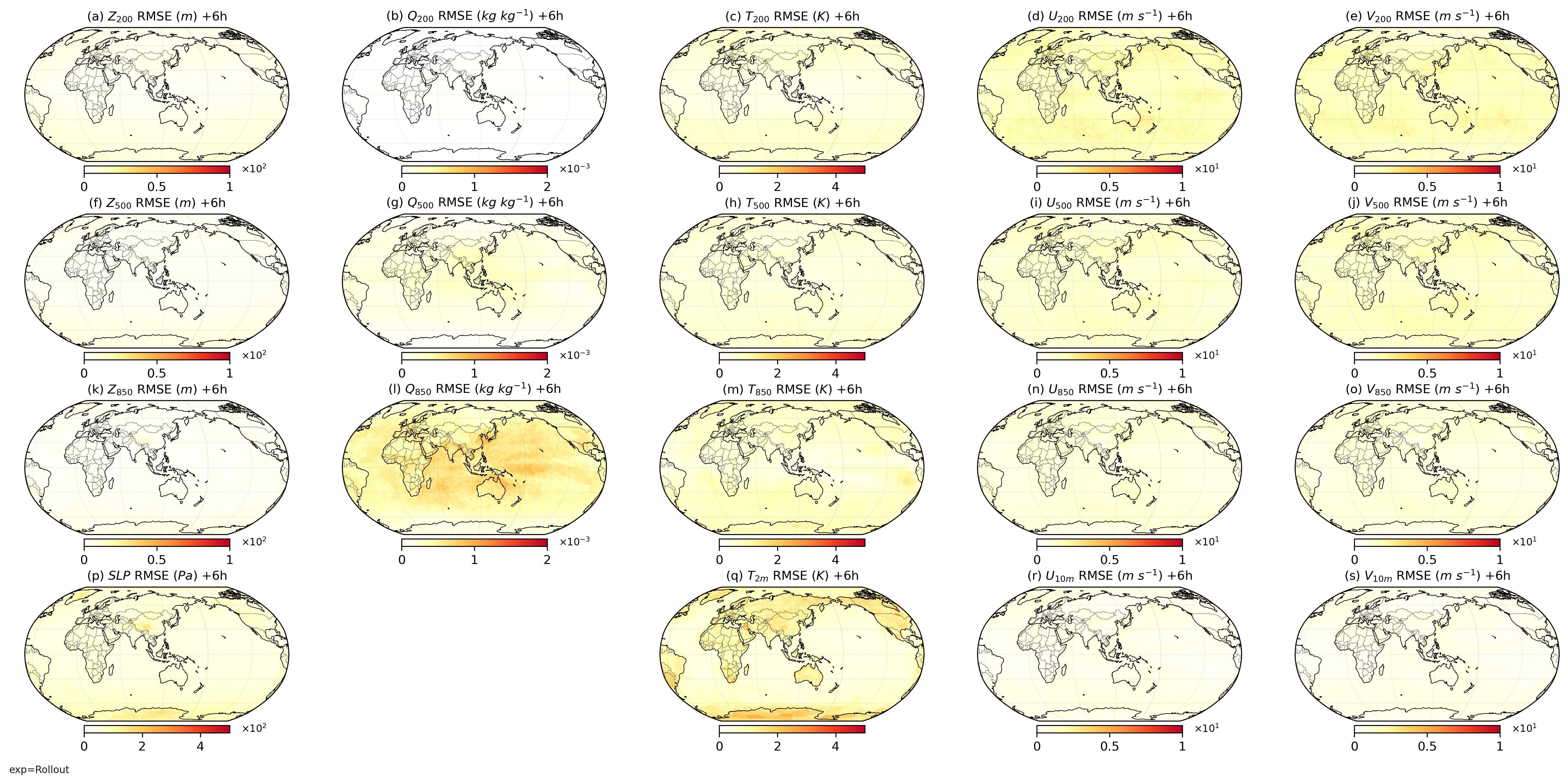}
\caption{Denormalized spatial \texttt{RMSE} distribution of the \texttt{SCS\_FCST4} model at \texttt{+6 h}. Panels (a-e) show \texttt{Z200}, \texttt{Q200}, \texttt{T200}, \texttt{U200}, and \texttt{V200}; panels (f-j) show \texttt{Z500}, \texttt{Q500}, \texttt{T500}, \texttt{U500}, and \texttt{V500}; panels (k-o) show \texttt{Z850}, \texttt{Q850}, \texttt{T850}, \texttt{U850}, and \texttt{V850}; panel (p) shows \texttt{SLP}; panel (q) shows \texttt{T2m}; panel (r) shows \texttt{U10m}; and panel (s) shows \texttt{V10m}. The color bar below each panel gives the \texttt{RMSE} magnitude in the corresponding physical unit.}
\label{fig:s5}
\end{figure}

\begin{figure}[H]
\centering
\includegraphics[width=\textwidth]{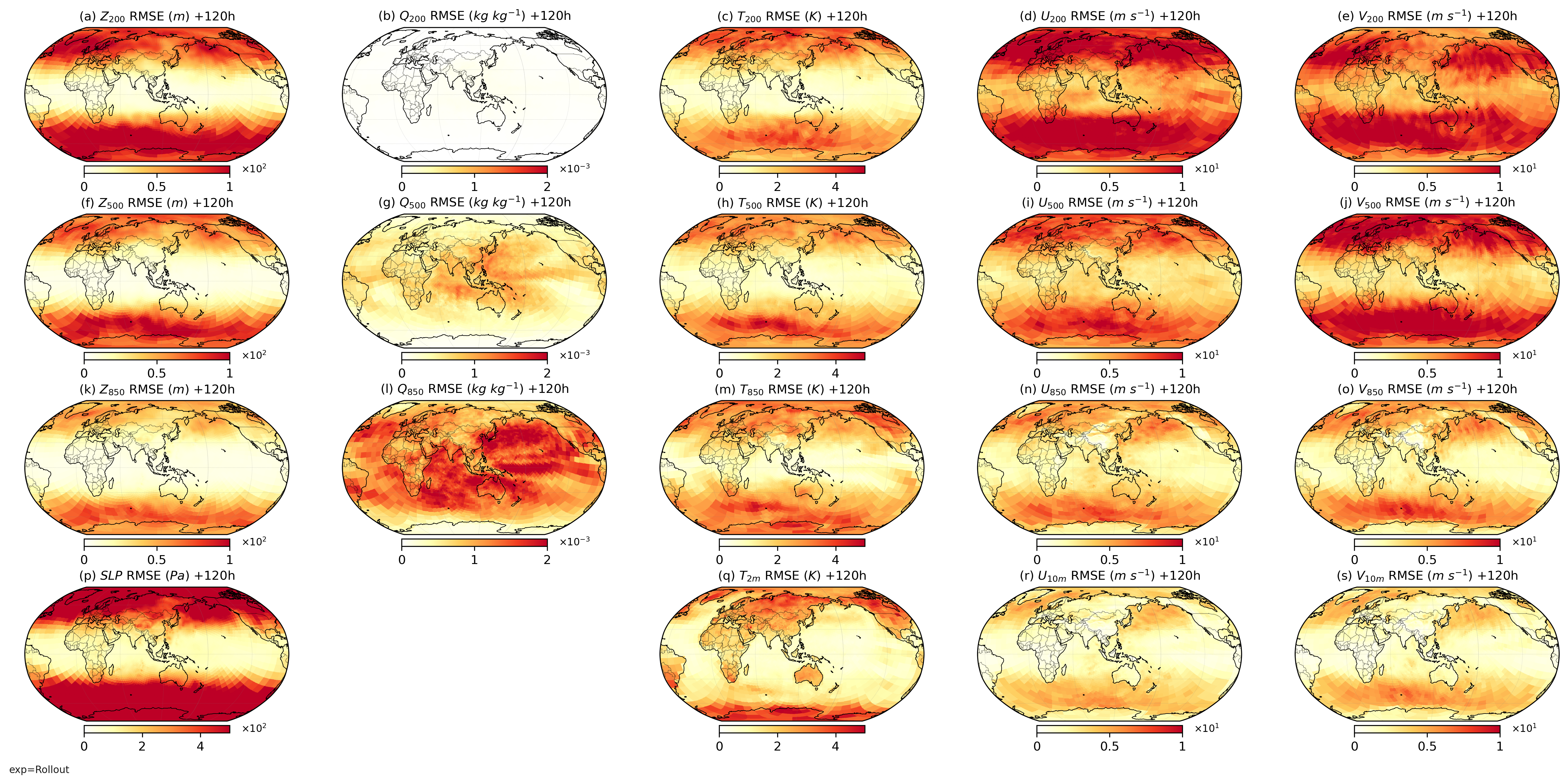}
\caption{Denormalized spatial \texttt{RMSE} distribution of the \texttt{SCS\_FCST4} model at \texttt{+120 h}. Panels (a-e) show \texttt{Z200}, \texttt{Q200}, \texttt{T200}, \texttt{U200}, and \texttt{V200}; panels (f-j) show \texttt{Z500}, \texttt{Q500}, \texttt{T500}, \texttt{U500}, and \texttt{V500}; panels (k-o) show \texttt{Z850}, \texttt{Q850}, \texttt{T850}, \texttt{U850}, and \texttt{V850}; panel (p) shows \texttt{SLP}; panel (q) shows \texttt{T2m}; panel (r) shows \texttt{U10m}; and panel (s) shows \texttt{V10m}. The color bar below each panel gives the \texttt{RMSE} magnitude in the corresponding physical unit.}
\label{fig:s6}
\end{figure}

This section supplements \seclink{sec:results-fcst4-avg}{5.1} of the main paper by expanding the averaged-performance analysis of \texttt{SCS\_FCST4 Model}. \figlink{fig:fig6}{6} in the main text shows the spatial \texttt{RMSE} distribution at \texttt{+36 h}, but a single lead time does not reveal whether the error structure is already diffuse from the beginning or accumulates gradually with time. \suppfiglink{fig:s5}{S5} and \suppfiglink{fig:s6}{S6} therefore add the \texttt{+6 h} and \texttt{+120 h} cases, allowing the reader to track how the same matrix of variables evolves from a short-range error baseline to medium-range error structure. Viewed together, the three figures show that error does grow, but the field does not quickly degenerate into structureless noise. It still preserves clear variable hierarchy and spatial organization, which means that the multistep model does not lose stability rapidly even in an averaged sense.

\begin{figure}[H]
\centering
\includegraphics[width=\textwidth]{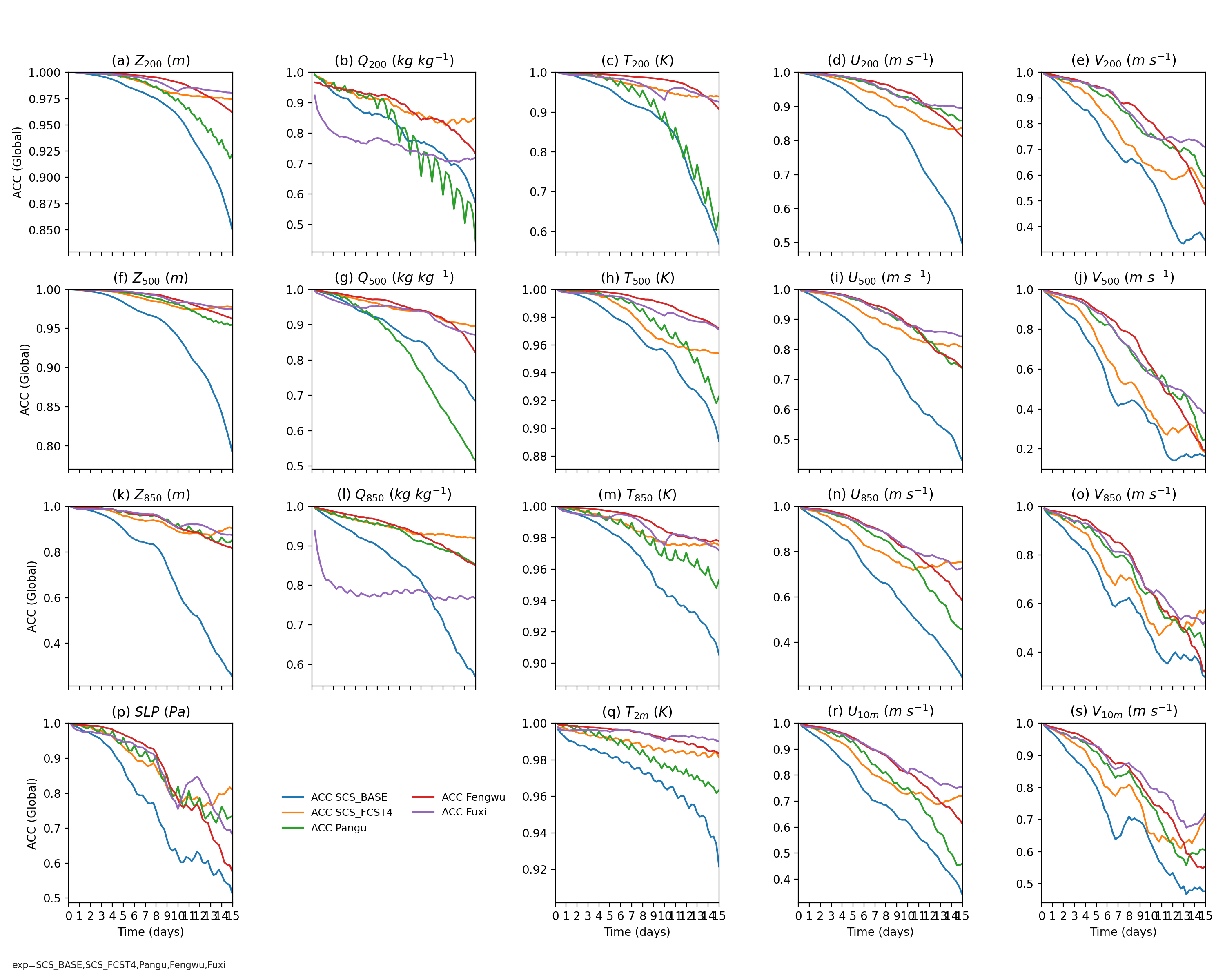}
\caption{Multi-model \texttt{ACC} time series over all six global faces. Panels (a-e) show \texttt{Z200}, \texttt{Q200}, \texttt{T200}, \texttt{U200}, and \texttt{V200}; panels (f-j) show \texttt{Z500}, \texttt{Q500}, \texttt{T500}, \texttt{U500}, and \texttt{V500}; panels (k-o) show \texttt{Z850}, \texttt{Q850}, \texttt{T850}, \texttt{U850}, and \texttt{V850}; panel (p) shows \texttt{SLP}; panel (q) shows \texttt{T2m}; panel (r) shows \texttt{U10m}; and panel (s) shows \texttt{V10m}. The x-axis is time in days and the y-axis is global \texttt{ACC}. Blue, orange, green, red, and purple curves denote \texttt{SCS\_Base}, \texttt{SCS\_FCST4}, \texttt{Pangu}, \texttt{FengWu}, and \texttt{FuXi}, respectively, and the legend is placed in the blank panel on the bottom row.}
\label{fig:s7}
\end{figure}

\figlink{fig:fig7}{7} in the main text focuses on the center-refined face because that is the core application region of the study. At the same time, the region outside the central face still provides global background constraints, so global evaluation also needs to be checked independently. \suppfiglink{fig:s7}{S7} extends the same \texttt{ACC} comparison to all \texttt{6} faces. The result should not be interpreted as a strict like-for-like competition against uniform global \texttt{0.25 degree} models, because the resolution outside the central region is much coarser. It is better interpreted as answering a different question: does center refinement in the present eastern-China configuration damage the consistency of the global background field? From that angle, \texttt{SCS\_FCST4} narrows the gap to the main comparison models substantially relative to \texttt{SCS\_Base}, and performs strongly on multiple variables. The benefit of rollout-oriented multistep training therefore remains visible not only in the central region, but also in the consistency of the overall circulation.

\subsection*{S4. Additional results for case studies of SCS\_FCST4}
\begin{figure}[H]
\centering
\includegraphics[width=\textwidth]{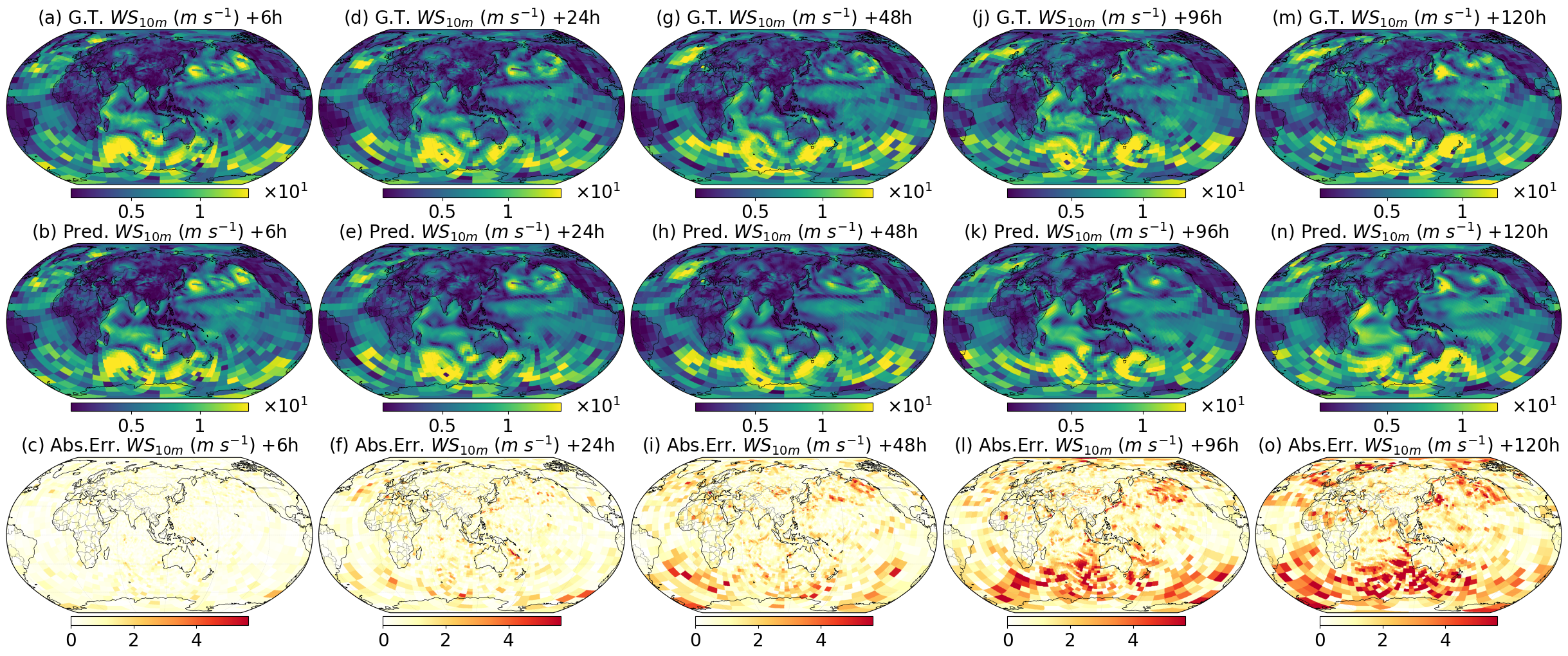}
\caption{Forecast of \texttt{10 m} wind speed (\texttt{WS10m}) for a representative case from the \texttt{SCS\_FCST4} model. Panels (a-e) show the ground truth (\texttt{GT}) at \texttt{+6 h}, \texttt{+24 h}, \texttt{+48 h}, \texttt{+96 h}, and \texttt{+120 h}; panels (f-j) show the prediction (\texttt{Pred}) at the same lead times; and panels (k-o) show the absolute error (\texttt{Abs. Err.}) at the same lead times. The first two rows share the same field color scale, the third row uses an absolute-error color scale, and the text in the lower-left corner reports the experiment name and sample index.}
\label{fig:s8}
\end{figure}

\begin{figure}[H]
\centering
\includegraphics[width=\textwidth]{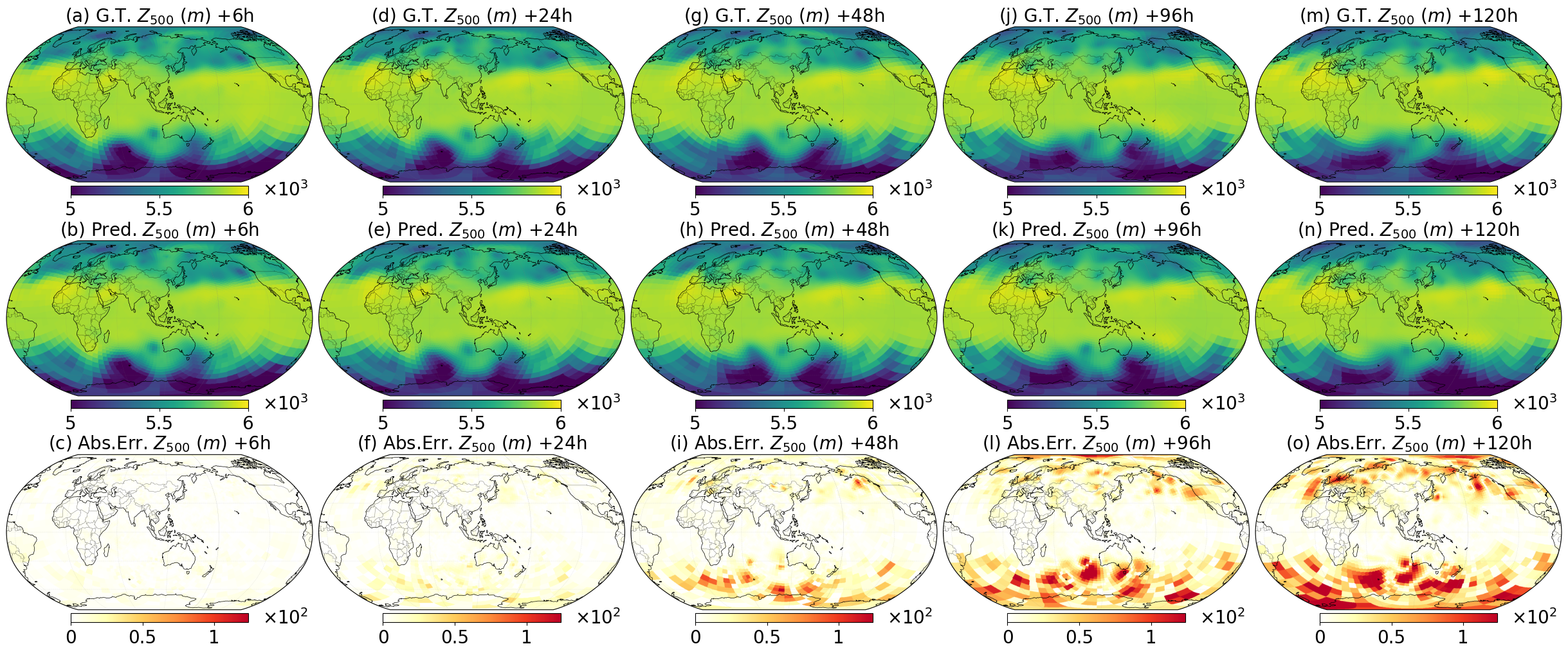}
\caption{Forecast of \texttt{500 hPa} geopotential height (\texttt{Z500}) for a representative case from the \texttt{SCS\_FCST4} model. Panels (a-e) show the ground truth (\texttt{GT}) at \texttt{+6 h}, \texttt{+24 h}, \texttt{+48 h}, \texttt{+96 h}, and \texttt{+120 h}; panels (f-j) show the prediction (\texttt{Pred}) at the same lead times; and panels (k-o) show the absolute error (\texttt{Abs. Err.}) at the same lead times. The first two rows share the same field color scale, the third row uses an absolute-error color scale, and the text in the lower-left corner reports the experiment name and sample index.}
\label{fig:s9}
\end{figure}

This section supplements \seclink{sec:results-fcst4-case}{5.2} of the main paper by extending the case-based analysis of \texttt{SCS\_FCST4 Model} across variables. \figlink{fig:fig8}{8} in the main text focuses on \texttt{Q850} and emphasizes the improved temporal continuity of low-level humidity after multistep training. That diagnosis is already enough to show better preservation of persistent moist regions and long-range moisture transport, but it does not answer whether the same improvement remains confined to humidity or also appears in other physically relevant fields. \suppfiglink{fig:s8}{S8} and \suppfiglink{fig:s9}{S9} therefore extend the same initialization time and the same lead times to \texttt{WS10m} and \texttt{Z500}, examining the gain from multistep training in the near-surface wind field and the midlevel circulation.

Viewed together, these three representative-case figures suggest that the improvement of \texttt{SCS\_FCST4} over \texttt{SCS\_Base} is not confined to a single variable, but instead reflects a broader contraction of error over the full forecast evolution. The elongated moisture-transport structures remain more complete, the strong-wind regions and wind-belt movement in the near-surface field stay more stable, and the trough-ridge positions and local error organization in \texttt{Z500} are also better controlled. Because these gains appear simultaneously across different physical layers, the main text can reasonably interpret the benefit of \texttt{SCS\_FCST4} as a systematic improvement in medium-range evolution under rollout-oriented training rather than an isolated gain in one class of variables.

\subsection*{S5. Long-range spectral diagnostics}
\begin{figure}[H]
\centering
\includegraphics[width=\textwidth]{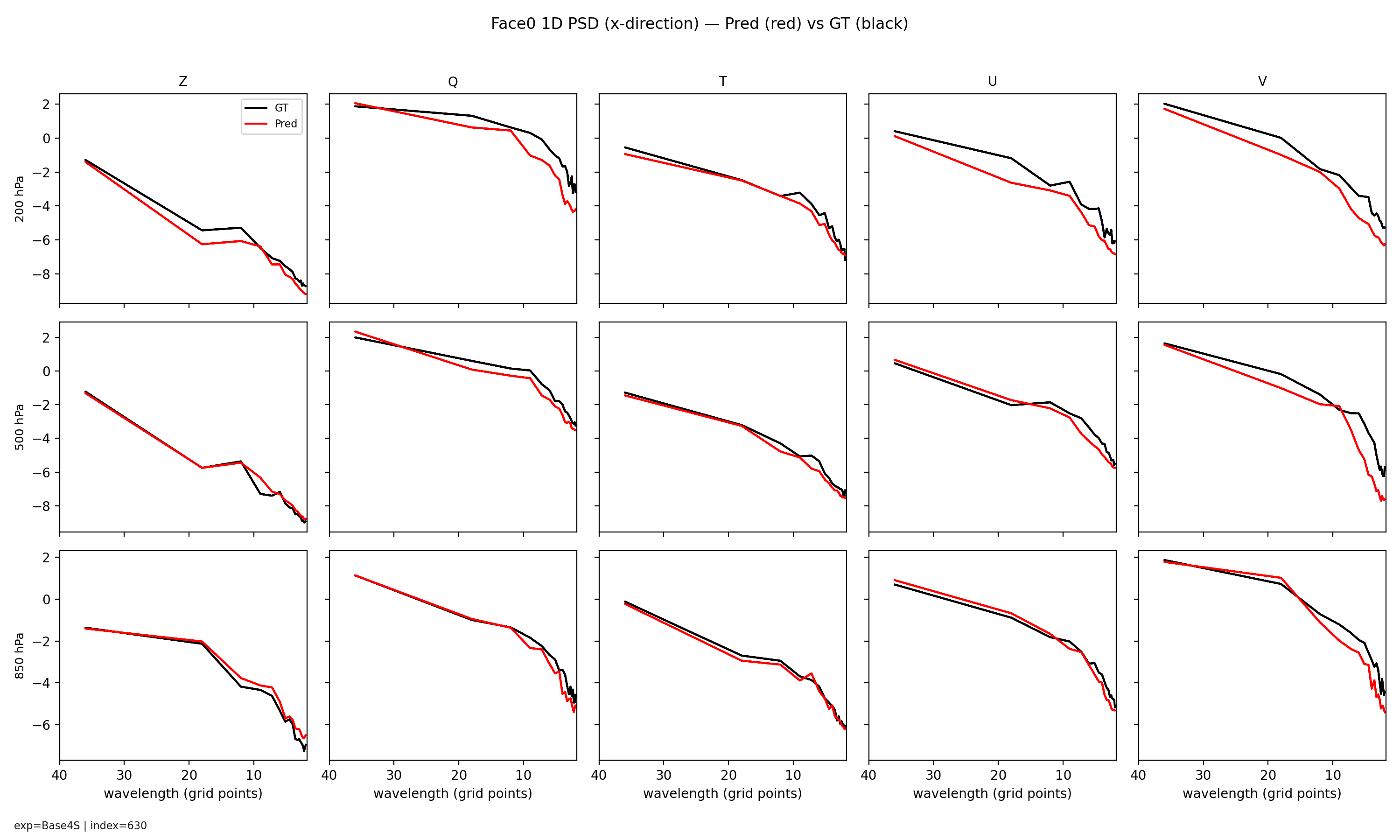}
\caption{Power spectral density (\texttt{PSD}) comparison on the center-refined face for the \texttt{SCS\_FCST4} model at \texttt{+120 h}. Columns correspond to \texttt{Z}, \texttt{Q}, \texttt{T}, \texttt{U}, and \texttt{V}, and rows correspond to \texttt{200 hPa}, \texttt{500 hPa}, and \texttt{850 hPa}. In each panel, the black curve denotes the ground truth (\texttt{GT}) and the red curve denotes the prediction (\texttt{Pred}); the x-axis is wavelength in grid points and the y-axis is log spectral energy.}
\label{fig:s10}
\end{figure}

\begin{figure}[H]
\centering
\includegraphics[width=\textwidth]{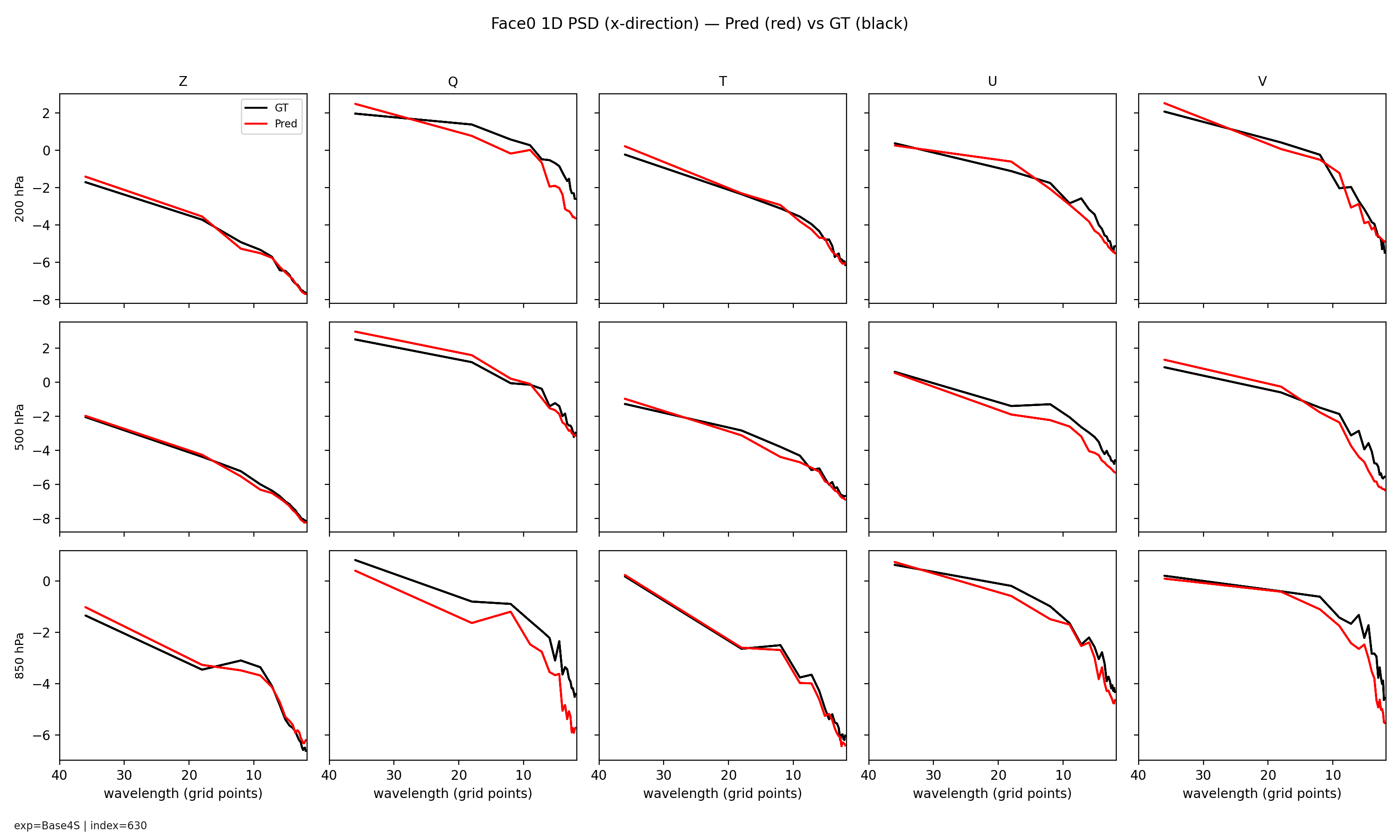}
\caption{Power spectral density (\texttt{PSD}) comparison on the center-refined face for the \texttt{SCS\_FCST4} model at \texttt{+240 h}. Columns correspond to \texttt{Z}, \texttt{Q}, \texttt{T}, \texttt{U}, and \texttt{V}, and rows correspond to \texttt{200 hPa}, \texttt{500 hPa}, and \texttt{850 hPa}. In each panel, the black curve denotes the ground truth (\texttt{GT}) and the red curve denotes the prediction (\texttt{Pred}); the x-axis is wavelength in grid points and the y-axis is log spectral energy.}
\label{fig:s11}
\end{figure}

This section supplements \seclink{sec:results-fcst4-psd}{5.4} of the main paper by extending the spectral diagnosis from \texttt{+36 h} to \texttt{+120 h} and \texttt{+240 h}. \figlink{fig:fig10}{10} in the main text already showed that \texttt{SCS\_FCST4} does not exhibit obvious collapse of small- and medium-scale energy at \texttt{+36 h}, especially in \texttt{Z} and \texttt{T}, where the spectrum remains comparatively stable from synoptic to mesoscale structure. The main question at longer lead times is whether that conclusion fails quickly as rollout continues. \suppfiglink{fig:s10}{S10} therefore shows the power spectral density at \texttt{+120 h}, focusing on whether the energy hierarchy from large to medium scales remains intact in the medium range. \suppfiglink{fig:s11}{S11} extends the lead time to \texttt{+240 h}, making it possible to inspect directly whether the spectral tail collapses monotonically and whether the degradation of different variables occurs in a synchronized way.

Viewed together, the key message of these figures is not that the long-range spectra remain identical to the ground truth. The important point is that spectral bias grows with lead time without turning into full instability. Geopotential height and temperature remain the most stable variables. Humidity shows a clearer lack of shortwave energy, and wind remains the hardest variable family, but the dominant spectral slope and the overall scale hierarchy of most variables still stay within a physically reasonable range. This supports the interpretation in the main text that the benefit of \texttt{SCS\_FCST4} is not limited to lower average error. Multistep training also suppresses the kind of long-range oversmoothing that is common in purely data-driven forecasting systems.

\begin{thebibliography}{25}
\bibitem[Bi et~al.(2023)]{bi2023pangu}
Bi, K., Xie, L., Zhang, H., Chen, X., Gu, X., \& Tian, Q. (2023).
\emph{Accurate medium-range global weather forecasting with 3D neural networks}.
Nature, 619, 533-538. \url{https://doi.org/10.1038/s41586-023-06185-3}

\bibitem[Bodnar et~al.(2025)]{bodnar2025aurora}
Bodnar, C., Bruinsma, W. P., Lucic, A., Stanley, M., Allen, A., Brandstetter, J., et al. (2025).
\emph{A foundation model for the Earth system}. Nature, 641, 1180-1187.
\url{https://doi.org/10.1038/s41586-025-09005-y}

\bibitem[Chen et~al.(2023)]{chen2023fuxi}
Chen, L., Zhong, X., Zhang, F., Cheng, Y., Xu, Y., Qi, Y., \& Li, H. (2023).
\emph{FuXi: A cascade machine learning forecasting system for 15-day global weather forecast}.
npj Climate and Atmospheric Science, 6, 190.
\url{https://doi.org/10.1038/s41612-023-00512-1}

\bibitem[Durran and Gingrich(2014)]{durran2014predictability}
Durran, D. R., \& Gingrich, M. (2014).
\emph{Atmospheric predictability: Why butterflies are not of practical importance}.
Journal of the Atmospheric Sciences, 71(7), 2476-2488.

\bibitem[Fox-Rabinovitz et~al.(2008)]{foxrabinovitz2008stretched}
Fox-Rabinovitz, M., Cote, J., Dugas, B., Deque, M., McGregor, J. L., \& Belochitski, A. (2008).
\emph{Stretched-grid Model Intercomparison Project: decadal regional climate simulations with enhanced variable and uniform-resolution GCMs}.
Meteorology and Atmospheric Physics, 100(1-4).

\bibitem[Gao et~al.(2025)]{gao2025oneforecast}
Gao, Y., Wu, H., Shu, R., Dong, H., Xu, F., Chen, R. R., Yan, Y., et al. (2025).
\emph{OneForecast: A Universal Framework for Global and Regional Weather Forecasting}.
arXiv:2502.00338v4. \url{https://arxiv.org/abs/2502.00338}

\bibitem[Harris and Lin(2013)]{harris2013twoway}
Harris, L. M., \& Lin, S.-J. (2013).
\emph{A Two-Way Nested Global-Regional Dynamical Core on the Cubed-Sphere Grid}.
Monthly Weather Review, 141(1), 283-306.
\url{https://doi.org/10.1175/MWR-D-11-00201.1}

\bibitem[Harris et~al.(2016)]{harris2016highresolution}
Harris, L. M., Lin, S.-J., \& Tu, C.-Y. (2016).
\emph{High-Resolution Climate Simulations Using GFDL HiRAM with a Stretched Global Grid}.
Journal of Climate, 29(12), 4293-4314.
\url{https://doi.org/10.1175/JCLI-D-15-0389.1}

\bibitem[Hersbach et~al.(2020)]{hersbach2020era5}
Hersbach, H., Bell, B., Berrisford, P., Hirahara, S., Horanyi, A., Munoz-Sabater, J., et al. (2020).
\emph{The ERA5 global reanalysis}. Quarterly Journal of the Royal Meteorological Society, 146(730), 1999-2049.
\url{https://doi.org/10.1002/qj.3803}

\bibitem[IBTrACS(2022)]{ibtracs2022muifa}
IBTrACS. (2022). \emph{Typhoon MUIFA (2022247N26147)}.
\href{https://ncics.org/ibtracs/index.php?name=v04r01-2022247N26147}{IBTrACS archive page}

\bibitem[Ji et~al.(2024)]{ji2024leadsee}
Ji, W., Feng, J., Liu, Y., Qiu, Y., \& Gao, H. (2024).
\emph{Leadsee-Precip: A Deep Learning Diagnostic Model for Precipitation}.
arXiv:2411.12640. \url{https://arxiv.org/abs/2411.12640}

\bibitem[Lam et~al.(2023)]{lam2023graphcast}
Lam, R., Sanchez-Gonzalez, A., Willson, M., Wirnsberger, P., Fortunato, M., Alet, F., et al. (2023).
\emph{Learning skillful medium-range global weather forecasting}. Science, 382(6677), 1416-1421.
\url{https://doi.org/10.1126/science.adi2336}

\bibitem[Lauritzen et~al.(2014)]{lauritzen2014transport}
Lauritzen, P. H., Ullrich, P. A., Jablonowski, C., et al. (2014).
\emph{A standard test case suite for two-dimensional linear transport on the sphere: results from a collection of state-of-the-art schemes}.
Geoscientific Model Development, 7, 105-145.
\url{https://doi.org/10.5194/gmd-7-105-2014}

\bibitem[Li et~al.(2024)]{li2024moganet}
Li, S., Wang, Z., Liu, Z., Tan, C., Lin, H., Wu, D., Chen, Z., Zheng, J., \& Li, S. Z. (2024).
\emph{MogaNet: Multi-order gated aggregation network}. ICLR.
\url{https://openreview.net/forum?id=XhYWgjqCrV}

\bibitem[Li et~al.(2020)]{li2020mcv}
Li, X., Chen, C., Shen, X., \& Xiao, F. (2020).
\emph{Development of a unified high-order nonhydrostatic multi-moment constrained finite volume dynamical core: derivation of flux-form governing equations in the general curvilinear coordinate system}.
arXiv:2004.05779. \url{https://doi.org/10.48550/arXiv.2004.05779}

\bibitem[Luo et~al.(2025)]{luo2025dragonboat}
Luo, X., et al. (2025).
\emph{Multiscale Factors Driving Extreme Flooding in China's Pearl River Basin During the 2022 Dragon Boat Precipitation Season}.
Water, 17(7), 1013. \url{https://www.mdpi.com/2073-4441/17/7/1013}

\bibitem[Molinaro et~al.(2026)]{molinaro2026diffusion}
Molinaro, R., Siegenheim, N., Martin, H., Frey, M., Poulsen, N., Seitz, P., \& Gabler, M. V. (2026).
\emph{Universal Diffusion-Based Probabilistic Downscaling}.
arXiv:2602.11893v2. \url{https://arxiv.org/abs/2602.11893}

\bibitem[NVIDIA(2026)]{earth2studio_docs}
NVIDIA. (2026). \emph{Earth2Studio Documentation}.
\url{https://nvidia.github.io/earth2studio/}

\bibitem[Park et~al.(2026)]{park2026srweather}
Park, H., Park, S., Kang, D., et al. (2026).
\emph{A super-resolution framework for downscaling machine learning weather prediction toward 1-km air temperature}.
npj Climate and Atmospheric Science, 9, 56.
\url{https://doi.org/10.1038/s41612-026-01328-5}

\bibitem[Perez et~al.(2018)]{perez2018film}
Perez, E., Strub, F., de Vries, H., Dumoulin, V., \& Courville, A. (2018).
\emph{FiLM: Visual reasoning with a general conditioning layer}.
Proceedings of the AAAI Conference on Artificial Intelligence, 32(1), 3942-3951.
\url{https://doi.org/10.1609/aaai.v32i1.11671}

\bibitem[Putman and Lin(2007)]{putman2007finite}
Putman, W. M., \& Lin, S.-J. (2007).
\emph{Finite-volume transport on various cubed-sphere grids}.
Journal of Computational Physics, 227(1), 55-78.
\url{https://doi.org/10.1016/j.jcp.2007.07.022}

\bibitem[RAMMB/CIRA(2022)]{rammb2022muifa}
RAMMB/CIRA. (2022). \emph{WP142022 - Typhoon Muifa archive}.
\url{https://rammb-data.cira.colostate.edu/tc_realtime/archive_text.asp?product=ripastbl&storm_identifier=wp142022}

\bibitem[Ren et~al.(2023)]{ren2023fengwu}
Ren, B., Liu, X., Li, S., et al. (2023).
\emph{FengWu: Pushing the skillful global medium-range weather forecast beyond the SOTA}.
arXiv:2304.02948. \url{https://arxiv.org/abs/2304.02948}

\bibitem[Xu et~al.(2025)]{xu2025yinglong}
Xu, P., Zheng, X., Gao, T., Wang, Y., Yin, J., Zhang, J., Zhang, X., et al. (2025).
\emph{An artificial intelligence-based limited area model for forecasting of surface meteorological variables}.
Communications Earth \& Environment, 6, 372.
\url{https://doi.org/10.1038/s43247-025-02347-5}

\bibitem[CW3E(2022)]{cw3e_eventsummary_june2022}
CW3E. (2022, June 15). \emph{CW3E Event Summary: 9-12 June 2022}.
Center for Western Weather and Water Extremes.
\url{https://cw3e.ucsd.edu/cw3e-event-summary-9-12-june-2022/}
\end{thebibliography}
\end{document}